\documentclass[pra, twocolumn, superscriptaddress]{revtex4-2}
\usepackage{graphicx,bm}
\usepackage{epsfig}
\usepackage{amsfonts}
\usepackage{amsmath}
\usepackage{amssymb}
\usepackage[caption=false]{subfig}
\usepackage{dsfont}
\usepackage{amsthm}
\usepackage{color}
\usepackage{mathrsfs}
\usepackage{enumitem}
\usepackage[colorlinks=true,linkcolor=blue,citecolor=blue,urlcolor=blue]{hyperref}
\usepackage{comment}
\usepackage{braket}
\usepackage{physics}
\usepackage{xr}
\usepackage[dvipsnames]{xcolor}
\usepackage{soul}
\usepackage[normalem]{ulem}
\usepackage{esvect}
\usepackage[T1]{fontenc}
\usepackage[utf8]{inputenc}

\definecolor{dgreen}{rgb}{0.0, 0.6, 0.0}

\newcommand{\D}{\mathrm{d}}
\newcommand{\argp}[1]{( #1 )}
\newcommand{\args}[1]{[ #1 ]}
\newcommand{\argc}[1]{\left\{ #1 \right\}}
\newcommand{\Abs}[1]{\left\vert #1 \right\vert}
\newcommand{\av}[1]{\left\langle #1 \right\rangle}

\DeclareFontEncoding{LS1}{}{}
\DeclareFontSubstitution{LS1}{stix}{m}{n}
\DeclareSymbolFont{stixletters}{LS1}{stix}{m}{it}
\DeclareMathAccent{\cev}{\mathord}{stixletters}{"91}

\definecolor{dgreen}{rgb}{0.0, 0.42, 0.24}

\NewDocumentCommand{\hj}{+m}{{\color{blue}[Jeong:#1]}}

\begin{document}

\title{How Quantum Is Light? \\
		 Concepts, Measures, and Applications of Optical Quantumness}

\author{Soumyakanti Bose}
\email{soumyakanti.bose09@gmail.com}
\affiliation{NextQuantum Center, Seoul National University, Seoul 08826, Korea}
\affiliation{Department of Physics, SRM University-AP, Amaravati, Andhra Pradesh 522240, India}

\author{Hyukjoon Kwon}
\email{hjkwon@kias.re.kr}
\affiliation{School of Computational Sciences, Korea Institute for Advanced Study, Seoul 02455, South Korea}

\author{Yong Siah Teo}
\email{ys\_teo@sejong.ac.kr}
\affiliation{Department of Quantum Information Science and Engineering, Sejong University, Seoul 05006, Republic of Korea}
\affiliation{NextQuantum Center, Seoul National University, Seoul 08826, Korea}

\author{Hyunseok Jeong}
\email{h.jeong37@gmail.com}
\affiliation{NextQuantum Center, Seoul National University, Seoul 08826, Korea}

\author{Luis L. S\'anchez-Soto}
\email{lsanchez@ucm.es}
\affiliation{Departamento de \'Optica, Facultad de F\'{\i}sica, Universidad Complutense, 28040~Madrid, Spain}
\affiliation{Institute for Quantum Studies, Chapman University, Orange, CA 92866, USA}
\affiliation{Max-Planck-Institut f\"ur die Physik des Lichts, 91058 Erlangen, Germany}


\begin{abstract}
Quantum states are ultimately responsible for numerous counterintuitive phenomena that defy classical intuition. 
As the development of quantum technologies accelerates, our ability to harness these nonclassical/quantum resources becomes essential. 
This review addresses a fundamental question: How do we quantify the ``quantumness'' of a state, and which specific traits distinguish it from its classical counterparts? 
We examine quantumness through both theoretical and operational lenses, providing a clear framework for state characterization. 
Ultimately, we demonstrate that certifying quantumness is not just a theoretical exercise; it is a practical necessity for identifying—and realizing—quantum advantage in real-world applications.
\end{abstract}

\maketitle

\section{Introduction}
\label{sec:introduction}

Quantum states of light are the cornerstone of modern information processing and offer deep insights into the true nature of the microscopic world~\cite{Pathak2017}. 
This naturally leads to the fundamental questions of how quantum is a state of light? 
What makes one state more quantum than another? 
Is there any trait to which one can point and say ``this embodies quantumness''?

Conceptually, our understanding of quantumness---a trait that no classical light can possess---is rooted in the Glauber-Sudarshan $P$ representation~\cite{Glauber1963, Sudarshan1963}, which allows us to expand each quantum state  $\rho$ into the coherent state basis.  
Because coherent states are the closest quantum versions of steady, classical waves, we would expect this $P$ representation to behave like a standard probability map in a classical world. 
However, since coherent states are not perfectly distinct from one another, this map does not always follow the rules of everyday logic.
A state reveals its nonclassical/quantum nature when the $P$~function takes negative values or becomes singular (to be precise, more singular than a Dirac delta function).
Since one cannot have a negative chance of something occurring in classical physics, these anomalies are the definitive proof that a state is behaving in a purely quantum manner.
  
However, because these singularities obstruct direct experimental measurement, the quest to quantify the nonclassicality has birthed a vast landscape of theoretical and empirical signatures. 
From foundational explorations in quantum theory to practical applications such as entanglement distribution, quantum cryptography, quantum computation, and quantum metrology, optical nonclassicality appears ubiquitously. 
This widespread relevance underscores the need for a distinct, concise, and comprehensive review of the historical developments that have shaped our understanding of this exotic feature of optical~states.

A word of caution is in order.  
In the early days of quantum optics, the term ``nonclassicality'' was primarily used to highlight the quantum nature of a state---a terminology that persists even today---yet implicitly views the world through a classical lens.  
From an intrinsic quantum perspective, however, every state is fundamentally quantum. 
To honor this reality, we instead adopt the term quantumness throughout this paper, centering our analysis within the quantum framework rather than defining it by what it is not.

Over the past three decades, numerous monographs, books, and review articles~\cite{Dodonov2002, Book_Agarwal2012, Book_Andrews2015,Schnable2017, Fabre2020, Book_Meystre2021} have provided detailed accounts of the developments in optical quantumness and its applications in modern science and technology, reflecting the maturity of the field~ \cite{Book_Gerry2005, Book_Vogel2006, Book_Agarwal2012}. 
While offering substantial insights into contemporary developments, each of these works tends to focus on specific and significant aspects of the intriguing nature of quantum light---whether it be the conceptual foundations for understanding quantumness \cite{Review_Loudon1980, Review_Kolobov1996}, experimental advances in engineering such states \cite{Review_DellAnno2006}, their technological applications \cite{Review_Braunstein2005, Review_Andersen2010, Review_Adesso2014, Pathak2017, Review_Tan2019}, or a pedagogical introduction to the mathematical framework to analyze such states \cite{Franson2018, Book_Serafini2023}. 
Nonetheless, a coherent and unified account that traces the historical evolution of the theory of quantumness---particularly one that systematically summarizes the various measures of quantumness and their progression into state-of-the-art technological applications---requires a fresh and modern approach.

This review provides a comprehensive introduction to the mathematical formulation, quantification, and technological relevance of quantumness. 
Besides meticulously attempting to catalog the vast body of existing literature, we focus on fundamental concepts and their physical interpretations as well. 
We begin in Sec.~\ref{sec:history} with a historical overview of how our understanding of quantum optical states was unified by the work of Glauber and Sudarshan, followed by a discussion of the mathematical framework and the operations that induce quantumness in Sec.~\ref{sec:definition}. 
Section~\ref{sec:signature} outlines the specific signatures of quantum light as they appear in both theory and the laboratory, while Sec.~\ref{sec:measure} offers a detailed look at how we measure and quantify these states using generalized frameworks. 
Sec.~\ref{sec:application} further explores the practical applications of these exotic states in quantum information and foundational physics, followed by several open issues as discussed in Sec. \ref{sec:openissues}. 
Finally, Sec. ~\ref{sec:conclusion} concludes with the summary.

\section{The notion of quantumness - a historical detour}
\label{sec:history}
\noindent 
{\bf 1930-1950: A mathematical endeavor to define phase-space distributions---} While a systematic and widely accepted description of deviations from classical electromagnetic theory emerged through the seminal works of Glauber~\cite{Glauber1963} and Sudarshan~\cite{Sudarshan1963}, the conceptual seeds of quantumness were planted much earlier, by Wigner in 1932. 
Shortly after Schrödinger formulated wave mechanics~\cite{Schrodinger1926} and Heisenberg established the uncertainty relation---restricting the simultaneous specification of canonically conjugate observables such as position and momentum \cite{Heisenberg1927, *Heisenberg_Book1930}---Wigner proposed an alternative formulation of quantum mechanics in phase space. 
This construction laid the groundwork for a quasi-probabilistic description of quantum systems.

In contrast to the Weyl correspondence~\cite{Weyl1927, Stone1936, Weyl_Book1950}, which associates classical phase-space functions with Hilbert-space operators, Wigner proposed an inverse mapping that assigns a phase-space function to a given quantum state. 
For an $n$-particle system, the Wigner function is defined as \cite{Wigner1932}
\begin{equation}
	W(\mathbf{x},\mathbf{p})  = \int \frac{\D^n\mathbf{x}^{\prime}}{(2 \pi)^{n}}~\Psi^*\argp{ \mathbf{x} +\tfrac{\hbar}{2}  \mathbf{x}^{\prime}} ~\Psi\argp{\mathbf{x} -\tfrac{\hbar}{2}\mathbf{x}^{\prime}}~e^{\mathsf{i} \mathbf{x}^{\prime} \cdot \mathbf{p}} \, ,
	\label{eq:wigner_nparticle}
\end{equation}
where $\mathbf{x}, \mathbf{p} \in \mathbb{R}^{n} $  are position and momentum, respectively.  

Despite its apparent departure from the standard Hilbert-space formulation~\cite{Dirac1930}, Wigner’s construction proved to be remarkably powerful.  Its formal consistency was established independently by Groenewold~\cite{Groenewold1946} and Moyal~ \cite{Moyal1949}, who developed a systematic phase-space formulation. 
Drawing on earlier insights by von Neumann~\cite{Neumann1931} and Dirac~\cite{Dirac1930a, *Dirac1945}, Groenewold~\cite{Groenewold1946} introduced a deformation of the classical Poisson bracket into the quantum commutator via the $\star$ product,
\begin{equation}
	(f \star g)(x,p) = f(x,p)~e^{\frac{\mathsf{i}\hbar}{2}   \overleftrightarrow{\mathcal{P}}}~ g(x,p) ,
\end{equation}
where $\overleftrightarrow{\mathcal{P}}$ is the Poisson operator
	\begin{equation}
		\label{eq:Poiss}
		\overleftrightarrow{\mathcal{P}} = 
		\frac{\overleftarrow{\partial}}{\partial x}
		\frac{\overrightarrow{\partial}}{\partial p} -
		\frac{\overleftarrow{\partial}}{\partial p}
		\frac{\overrightarrow{\partial}}{\partial x} \, .
	\end{equation}
Within this formalism, the Wigner function naturally emerges as the phase-space representation of the density operator.

For a single-particle system, Eq.~\eqref{eq:wigner_nparticle} reduces to
\begin{align}
	W(x,p) & = \int \frac{\D x^{\prime}}{2 \pi}~\Psi^*\argp{x+\tfrac{\hbar}{2}x^\prime} ~\Psi\argp{x-\tfrac{\hbar}{2} x^\prime} ~e^{\mathsf{i} x^\prime p}
	\nonumber 
	\\
	& = \int \frac{\D x^{\prime}}{2 \pi}~\Psi(x)\star e^{\mathsf{i} x^\prime p}\star\Psi^*(x),
	\label{eq:wigner_star}
\end{align}
enabling a direct connection between operator algebra in Hilbert space and function algebra in phase space. 
This formulation admits a natural generalization to arbitrary operators and statistical ensembles, as further developed by Moyal \cite{Moyal1949}. 
For a comprehensive treatment, see \cite{Zachos2005}.

It was later established that the Wigner function is bounded~\cite{Baker1958}
\begin{equation}
	-\frac{2}{\hbar} \leq W(x,p) \leq \frac{2}{\hbar},
\end{equation}
which further underscores its quasiprobabilistic nature.

These developments also resolved a conceptual tension lurking at the heart of the phase-space picture: how can a distribution defined over both position and momentum coexist with Heisenberg's uncertainty relation? 
The resolution lies in the inherent ``fuzziness'' of the Wigner function. 
Although it ranges over all of phase space, it cannot represent arbitrarily sharp localization---any effective concentration is constrained to regions of area on the order of $\hbar$. 
This built-in indeterminacy reflects the noncommutativity of conjugate observables and ensures that the uncertainty relation remains inviolate.
	
In parallel to efforts to formalize such quasiprobability distributions—often characterized by nonpositivity or singular behavior—there emerged an alternative approach aimed at representing quantum states using smooth, positive-definite phase-space functions. 
This direction was initiated by Husimi~\cite{Husimi1940}, who introduced a Gaussian-smoothed version of the Wigner function. 
A systematic construction of such distributions within an operator-ordering framework was provided much later by Kano~\cite{Kano1965}. 
Unlike the Wigner function, which can assume negative values and thus directly signal quantumness, these smoothed distributions are everywhere non-negative and free from singularities. 
A clear physical interpretation of these positive-definite phase-space distributions was established, further later, by Wehrl~\cite{Wehrl1978, Wehrl1979}. 
This provided an information-theoretic perspective on phase-space localization and further strengthened the connection between classical and quantum descriptions (we shall discuss later).

\noindent 
{\bf 1954-1963: A new ripple in optical coherence theory---}
Alongside the development of phase-space formulations based on classical-like distributions over conjugate variables, a separate line of investigation took shape that focused on the statistical properties of photons in cavity-quantized radiation fields. 
This direction was significantly influenced by the Hanbury Brown-Twiss (HBT) experiment~\cite{Brown1956}---which revealed unexpected intensity correlations in optical fields---later recognized as photon bunching.

Motivated originally by radioastronomical measurements of stellar angular diameters—most notably of Sirius \cite{Brown1956a}---Hanbury Brown and Twiss devised an optical interferometric setup to probe intensity correlations.  In this configuration, an incoming optical field is split by a half-silvered mirror (effectively a 50:50 beam splitter) into transmitted and reflected beams, each directed to its own photodetector.  The intensities are recorded independently, and the key observable is the second-order intensity correlation, obtained by coincidence counting of time-resolved detection events.

Formally, the HBT experiment measures the normalized second-order correlation function
\begin{equation}
	g^{(2)}(\tau) = \frac{\av{I(t)I(t+\tau)}}{\av{I(t)}^2},
    \label{eq:g2tau}
\end{equation}
which quantifies temporal correlations between photon detection events.
Classical light follows the conditions: (a) $g^{(2)}(0) > g^{(2)}(\tau)$ and $g^{(2)}(0) \geq 1$.
For example, thermal light yields $g^{(2)}(0) = 2$, indicating bunching, whereas deviations from this behavior---particularly $g^{(2)}(0) < 1$ and $g^{(2)}(0) < g^{(2)}(\tau)$---are the hallmarks of distinctly quantum radiation.
This is also known as the anti-bunching.

\begin{figure}
	\centering
	\includegraphics[scale=1]{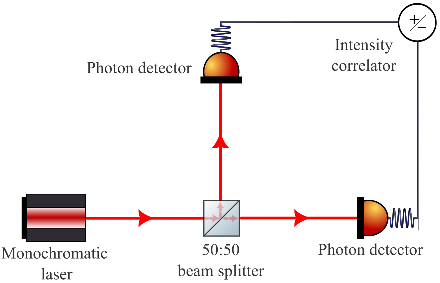}
	\caption{Schematic of Hanbury Brown--Twiss experiment where an incoming signal is split into a half-silvered mirror and then fed to two detectors registering the intensity.
		Subsequently the data is fed to a monitor to analyze intensity correlation between the detectors.}
	\label{fig:BrownTwiss}
\end{figure}

To avoid the technical complexities associated with the original radiofrequency detectors \cite{Brown1954}, we consider here an idealized version of the HBT setup, as illustrated in Fig.~\ref{fig:BrownTwiss}, which captures the essential physics of intensity correlation measurements.

While the results of the original radiofrequency experiments were fully consistent with Maxwell’s electromagnetic theory, the corresponding optical experiments gave rise to a genuine puzzle: 
the measured intensity correlation grew stronger as the separation between detectors shrank, pointing to surprisingly strong correlations in the detected light. 
Similar effects were reported in related experiments, including those by Rebka and Pound \cite{Rebka1957}.

A satisfactory semiclassical explanation of the HBT effect---as a statistical property of partially coherent light---was first provided independently by Purcell~\cite{Purcell1956}. 
This interpretation was subsequently refined by Hanbury Brown and Twiss themselves \cite{Brown1957, *Brown1958}, who analyzed the phenomenon in terms of intensity fluctuations in partially coherent sources. 
From a different angle, Mandel \cite{Mandel1959, *Mandel1961} attributed the observed enhancement in photon correlations, both in HBT experiments and maser systems, to the narrow spectral bandwidth of the radiation field.
Yet these largely semiclassical frameworks proved inadequate for a unified, systematic treatment of higher-order correlation functions~\cite{Born_Wolf_Book1959}---exposing the need for a genuinely quantum theory of optical coherence.

Glauber met this challenge: by building on his earlier work~\cite{Glauber1951} and related contributions by Fano~\cite{Fano1961}, he formulated a hierarchy of correlation functions and a consistent photodetection theory~\cite{Glauber1963a}---a framework that broke sharply with the semiclassical treatments that came before, including Mandel's. 
Glauber then extended this formalism~\cite{Glauber1963b} to characterize quantum states of light in terms of coherent-state representations~\cite{Glauber1963}. 
Within this framework, any density operator $\rho$ can be expressed as a statistical mixture of coherent states,
\begin{equation}
	\rho =  \int \frac{\D^2\alpha}{\pi}~P(\alpha) \ket\alpha\bra\alpha, 
	\label{eq:SGP_function}
\end{equation}   
where the \emph{weight function} $P(\alpha)$ might fail to maintain positivity hinting at a significant departure from the \emph{classical theory of coherence}. 

Complementary to Glauber’s formulation, Sudarshan~\cite{Sudarshan1963} sought a general representation of arbitrary quantum states as distributions over the overcomplete set of coherent states \cite{Klauder1960, *Bargmann1961, *Schweber1962}. 
While Glauber's approach was rooted in correlation functions, Sudarshan offered a more direct structural picture through the so-called diagonal representation, in which the $P(\alpha)$ in \eqref{eq:SGP_function} is a quasiprobability distribution over coherent states. 
Glauber had already noted that $P(\alpha)$ may fail to remain positive for quantum states; Sudarshan’s formulation placed this observation on a more rigorous footing by explicitly \emph{allowing $P(\alpha)$ to be a generalized function}. 
As a result, in the quantum regime, $P(\alpha)$ may exhibit strong singularities---such as derivatives of delta functions---reflecting the intrinsically nonclassical (quantum) nature of the underlying state. 
A detailed modern perspective on these aspects can be found in \cite{Franson2018}.

Importantly, in both Glauber’s and Sudarshan’s formulations, the diagonal coherent-state representation provides a natural criterion for quantumness: a state is classical if and only if $P(\alpha)$ behaves as a well-defined, positive semi-definite probability distribution. Any violation of this condition—negativity or singular behavior—signals quantumness. Moreover, this criterion admits a direct operational interpretation via photodetection theory, where normally ordered correlation functions and coincidence measurements~\cite{Kelley1964, Mandel1965, Luoisell_Book1973} can be expressed as averages over $P(\alpha)$.

\noindent
{\bf 1963-1970: A confluence of two approaches and emergence of a unified mathematically consistent theory---}Despite their distinct conceptual origins, the two approaches outlined above ultimately converge into a unified framework for the phase-space description of quantum states of light~\cite{Mehta1965, Mandel1965, Aharonov1966, Klauder_Sudarshan_Book1968, *Klauder_Sudarshan_Book2006}. 
This synthesis is rooted in the Wigner-Weyl correspondence, first formalized by Groenewold \cite{Groenewold1946}, and later recognized as a Fourier-transform-based mapping between Hilbert-space operators, quasiprobability distributions, and their associated characteristic functions~\cite{Segal1961}. 
These developments enabled a systematic generalization of phase-space methods for arbitrary operators~\cite{Mehta1967, Cahill1969}.

A particularly elegant outcome of this formalism is the introduction of the $s$-parametrized family of quasiprobability distributions by Cahill and Glauber~\cite{Cahill1969}, defined as
\begin{equation}
	f^{(s)}(z) = \int \frac{\D^2\lambda}{\pi}~ e^{s\frac{|\lambda|^2}{2}} \text{Tr} [\rho \hat{D}(\lambda)]~e^{z\lambda^* - z^*\lambda},
	\label{eq:spsd_genform}
\end{equation} 
where $\hat{D}(\lambda) = e^{\lambda \hat{a}^\dagger - \lambda^*\hat{a}}$ is the displacement operator, and the correspondence $z\leftrightarrow \hat{a}$ and $z^*\leftrightarrow \hat{a}^\dagger$ defines the phase-space variables, with $z= (x+\mathsf{i}p)/\sqrt{2}$. Recognizing $e^{z\lambda^* - z^*\lambda}$ as the Fourier kernel, one immediately obtains the associated $s$-ordered characteristic function,
\begin{equation}
	\chi^{(s)}(\lambda) = e^{s\frac{|\lambda|^2}{2}} \text{Tr} [\rho \hat{D}(\lambda)] ,
	\label{eq:scf_genform}
\end{equation}
which serves as the Fourier transform of the corresponding quasiprobability distribution.

Within this family, the Glauber-Sudarshan $P$-function, the Wigner function $W$, and the Husimi-Kano $Q$-function emerge as special cases corresponding to normal ($s=+1$), symmetric ($s=0$), and antinormal ($s=-1$) operator ordering, respectively.

More generally, the correspondence between Hilbert-space operators and phase-space functions can be formulated as an operator-valued Fourier transform ~\cite{Agarwal1970a, Agarwal1970b}. 
Given a phase-space function $F(z)$ associated with an operator $\hat{G}(\hat{a},\hat{a}^\dagger)$, one defines a linear mapping $\Omega$ such that $\Omega \args{ F(z) } = \hat{G}(\hat{a},\hat{a}^\dagger)$. 
This mapping can be expressed as
\begin{align}
	F(z) & = \int \frac{\D^2\alpha}{\pi}~e^{z\alpha^* - z^*\alpha}\, \text{Tr} [ \hat{G}(\hat{a},\hat{a}^\dagger) \hat{D}_\Omega(\alpha)],
	\label{eq:psd_opfourier}
\end{align}
where $D_\Omega(\alpha)$ encodes the operator-ordering rule (e.g., normal, symmetric, or antinormal ordering). In this sense, different quasiprobability distributions arise from different choices of ordering prescriptions~\cite{Agarwal1970a}.

\begin{figure}
	\centering
	\includegraphics[scale=0.45]{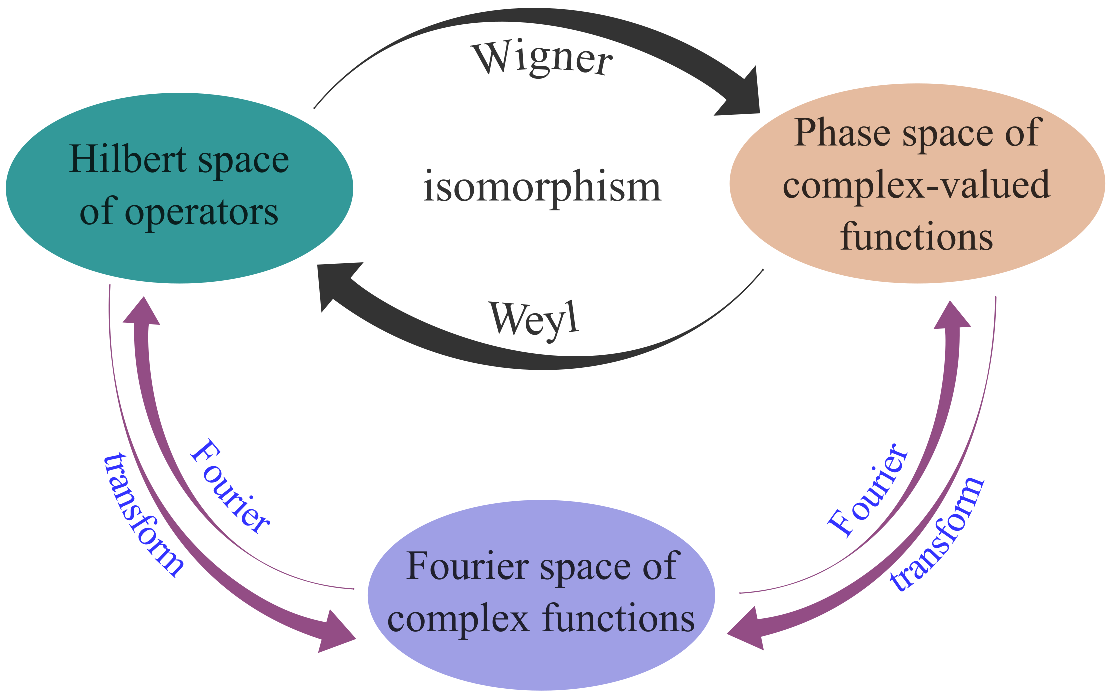}
	\caption{Schematic representation of Wigner-Weyl correspondence between Hilbert-space operators and Phase-space distributions through operator-Fourier transformation}
	\label{fig:WignerWeyl}
\end{figure}

A schematic representation of this Wigner–Weyl correspondence, highlighting the role of Fourier transformations in connecting operators and phase-space distributions, is shown in Fig.~\ref{fig:WignerWeyl}.

Applying this formalism to the density operator $\rho$, one obtains explicit expressions for the standard quasiprobability distributions \cite{Agarwal1970a}:
\begin{subequations}
	\begin{equation}
		P(z) = e^{|z|^2} \int \frac{\D^2\alpha}{\pi}~e^{|\alpha|^2} \bra{-\alpha} \rho \ket{\alpha} e^{z\alpha^* - z^*\alpha}
		\label{eq:Prho}
	\end{equation}
	\begin{equation}
		W(z)=  2e^{2|z|^2} \int \frac{\D^2\alpha}{\pi}~\bra{-\alpha} \rho \ket{\alpha} e^{2(z\alpha^* - z^*\alpha)}
		\label{eq:Wrho}
	\end{equation}
	\begin{equation}
		Q(z)= \bra{z} \rho \ket{z},
		\label{Qrho}
	\end{equation}
	\label{eq:psd_expression}
\end{subequations}
where $P$, $W$, and $Q$ correspond to normal, symmetric, and antinormal ordering, respectively.

These three distributions, however, are far from being interchangeable in character. 
The Glauber–Sudarshan $P$-function can become severely singular in the quantum regime—surviving only as a generalized function, sometimes involving derivatives of delta distributions—and may take negative values. 
The Wigner function, by contrast, is always a well-defined real function; it too can go negative, but remains bounded. 
The Husimi $Q$-function stands apart again: it is everywhere positive and bounded, satisfying $0\leq Q(z) \leq 1$. 
One might be tempted to read this as a genuine probability distribution—but that temptation should be resisted. 
Despite its regularity, the $Q$-function fails to reproduce the correct marginal distributions for conjugate variables~\cite{Takahashi1986, Takahashi1989}, unlike the Wigner function, which does recover the correct position and momentum marginals. 
In other words, \emph{its positivity is bought at the price of an intrinsic smoothing that blurs the quantum features.}

Nonetheless, these quasiprobability distributions are related to one another via Gaussian convolution relations,
\begin{subequations}
	\begin{equation}
		W(z) = 2 \int \frac{\D^2\alpha}{\pi}~P(\alpha) e^{-2|z-\alpha|^2}
		\label{eq:p2w}
	\end{equation}
	\begin{equation}
		Q(z) = \int \frac{\D^2\alpha}{\pi}~P(\alpha) e^{-|z-\alpha|^2}
		\label{eq:p2q}
	\end{equation}
	\begin{equation}
		Q(z) = 2 \int \frac{\D^2\alpha}{\pi}~W(\alpha) e^{-2|z-\alpha|^2},
		\label{eq:w2q}
	\end{equation}
	\label{eq:psd_interrelation}
\end{subequations}
which explicitly demonstrate that the $W$ and $Q$-functions arise as progressively smoothed versions of the $P$-function.
From this perspective, the hierarchy $P\rightarrow W \rightarrow Q$ reflects a gradual suppression of fine phase-space features, including those associated with quantumness.  In particular, such Gaussian smoothing provides an intuitive route to understanding the quantum-to-classical transition within the phase-space formalism~\cite{Leaf1968}. 

In what follows, we shall drop the conjugated variable in all phase-space functions for notational~simplification.

\noindent 
{\bf 1970-1985: Development of semiclassical theories and mathematical extension of Weyl transformations---} 
With the formal framework of phase-space methods now firmly established, the following years brought a wave of new developments along several fronts: 
deeper investigations into the quantum-to-classical transition, rigorous mathematical analyses of the Wigner-Weyl transform, statistical formulations of quantum mechanics, and extensions of the Weyl-Heisenberg algebra to finite-dimensional systems such as spin degrees of freedom.

In this review, however, we conclude the present section by highlighting selected developments most relevant to our discussion: semiclassical phase-space methods~\cite{Heller1977, Berry1977, Berry1977a, Berry1979} and generalizations of the Glauber–Sudarshan $P$-function~\cite{Drummond1980}. 
These efforts collectively contributed to a mature and unified understanding of phase-space distributions, as systematically reviewed in~\cite{Hillery1984}.
Taken together, these developments established a unified framework in which quantum states can be analyzed through classical-like phase-space distributions, differing primarily in their regularity, positivity, and sensitivity to quantum effects.

While substantial progress has been made toward a comprehensive phase-space description of quantum systems---as will be explored in the following sections---recent works \cite{Jeongwoo2017, Damanet2018, Sperling2018, Franson2018, Sperling2020} offer alternative perspectives that revisit and extend the conventional framework. 
These approaches provide fresh insights into the structure, interpretation, and quantification of the quantum optical states.

\section{Definition and few important results}
\label{sec:definition}

Let us consider a coherent state $\ket\alpha$, the closest physical realization of the output of a stabilized laser. 
Using the defining relation of the Glauber-Sudarshan $P$-function~\eqref{eq:Prho}, one finds 
\begin{align}
	P_{\ket\alpha}(z) &= \delta^2(z - \alpha) , 
	\label{eq:Palpha}
\end{align}
that is, a distribution localized at a single point in phase space.  \emph{This delta-function form of $P_{\ket\alpha}(z)$ does not imply that the state is sharply localized in phase space}; rather, it reflects the fact that coherent states form an overcomplete basis and behave as classical phase-space points within the $P$-representation.  In contrast, their Wigner and Husimi functions remain Gaussian distributions with finite width, consistent with the uncertainty relation.

A key property of coherent states is that they reproduce all normally ordered correlation functions of classical electromagnetic fields~\cite{Born_Wolf_Book1959}, which is why they are regarded as the most classical pure states of the radiation field—a statement that can be formalized rigorously~\cite{Hillery1985}. Since classical statistical mixing cannot itself generate quantumness, any convex combination of coherent states with a positive weight function remains classical. 
Accordingly, the most general classical state can be written as
\begin{equation}
	\rho_{\text{cl}} = \int \frac{\D^2\alpha}{\pi}~f_\text{cl}(\alpha) \ket\alpha\bra\alpha,
	\label{eq:def_classical}
\end{equation}
where $f_\text{cl}(\alpha) \geq 0$ is a well-defined probability distribution. 
The corresponding $P$-function is then simply
\begin{equation}
	P_\text{cl}(z) = \int \frac{\D^2\alpha}{\pi}~f_\text{cl}(\alpha)\,\delta^2 (z-\alpha),
\end{equation}
which is itself positive and regular.

Quantumness, in this language, emerges precisely when this picture breaks down—when $P$ ceases to behave as a genuine probability distribution, either by going negative or by developing singularities beyond simple delta functions. 
This raises a natural question: what operations drive a state across this boundary? There is no single answer, but several illuminating cases are worth examining.

\noindent
{\bf Photon-added states are quantum}---
Quantization of the electromagnetic field amounts to recasting each mode of the field as an effective quantum harmonic oscillator, with discrete energy levels separated by the fundamental quantum $\hbar\omega$. The corresponding excitations---photons---represent the elementary quanta of the field. 
A natural question, therefore, is whether the addition of such quanta necessarily induces quantumness.

To address this, consider an arbitrary initial state characterized by a Glauber–Sudarshan distribution $P_\text{ini}(\alpha)$, and define the $m$-photon–added state (PAS~\cite{Agarwal1991, Zavatta2004}) as
\begin{equation}
    \rho_\text{pas} = \frac{1}{N_m} \int \frac{\D^2\alpha}{\pi} \, P_\text{ini}(\alpha)\,\hat{a}^{\dagger m} \ket\alpha\bra\alpha \hat{a}^m,
    \label{eq:rho_pas}
\end{equation}
where $N_m$ ensures normalization.

Using the definition of the $P$-function, one finds that the corresponding phase-space representation takes the form
\begin{align}
    P_\text{pas}(z) &= \frac{e^{|z|^2}}{N_m} \int \frac{\D^2\beta}{\pi}\,  e^{|\beta|^2} \int\frac{d^2\alpha}{\pi} P_\text{ini}(\alpha) 
    \nonumber 
    \\
    &\qquad\qquad\times \bra{-\beta} \hat{a}^{\dagger m} \ket\alpha\bra\alpha \hat{a}^m\ket{\beta} e^{z\beta^* - z^*\beta}
    \nonumber 
    \\
    &= \frac{1}{N_m} \int \frac{\D^2\alpha}{\pi} P_\text{ini}(\alpha) e^{|z|^2 - |\alpha|^2} 
    \nonumber 
    \\
    &\qquad\qquad\times \partial_z^m \partial_{z^*}^m \int\frac{d^2\beta}{\pi} e^{(\alpha^* - z^*)\beta - (\alpha - z)\beta^*}
    \nonumber 
    \\
    &= \frac{e^{|z|^2}}{N_m } \int \dfrac{\D^2\alpha}{\pi}  P_\text{ini}(\alpha) e^{-|\alpha|^2} \partial_z^m \partial_{z^*}^m \delta^2(z - \alpha).
    \label{eq:Ppas}
\end{align}

This expression shows that photon addition transforms the initial $P$-distribution through derivatives of delta functions. 
However, this does not necessarily imply that the resulting $P$-distribution is singular.
For example, adding $n$ photons to the vacuum produces the Fock state $\left|n\right\rangle$ ($n\geq 1$), whose $P$-function is
\begin{equation}
    P_{\left|n\right\rangle}(z)
    = \frac{e^{|z|^2}}{n!}
    \partial_z^n\partial_{z^*}^n\delta^2(z).
    \label{eq:Pfock}
\end{equation}

This singular structure reflects the strongly quantum character of Fock states.
By contrast, photon-added thermal states can have smooth but negative $P$-functions~\cite{Kiesel2008}. 
Thus, photon addition provides a clear example of how optical quantumness can arise without necessarily producing a singular $P$-distribution.
Nevertherless, photon addition necessarily generates quantumness, with the resulting $P$-distribution either becoming negative or exhibiting singularities beyond those of ordinary delta functions.

\noindent
{\bf Photon subtraction induces no quantumness}---
In contrast to photon addition, one might expect photon subtraction to also act as a quantumness-inducing operation. 
However, this is not generally the case. To see this, consider the action of the annihilation operator on a coherent state, $\hat{a}\ket\alpha = \alpha\ket\alpha$. 
Using this property, a photon-subtracted state (PSS) constructed from an arbitrary initial state with $P$-function $P(\alpha)$ can be written as
\begin{equation}
	\rho_\text{pss} = \frac{1}{N_m} \int \dfrac{\D^2\alpha}{\pi}~P(\alpha)\,\hat{a}^{m} \ket\alpha\bra\alpha \hat{a}^{\dagger m} \, ,
	\label{eq:rho_pss}
\end{equation}
and the corresponding $P$-function is
\begin{equation}
	P_\text{pss}(z) = \frac{1}{N_m} \int \dfrac{\D^2\alpha}{\pi}~P(\alpha)\, |\alpha|^{2m} \delta^2(z - \alpha).
	\label{eq:Ppss}
	\end{equation}

The effect is simply to rescale the original distribution by the positive factor $\Abs{\alpha}^{2m}$—no new singularities, no new negativity. 
A classical input therefore yields a classical output: photon subtraction preserves whatever classicality (or quantumness) was already present, but introduces none of its own.

\noindent
{\bf Quantumness can be induced by quadrature squeezing}---Let us now consider a classical state $\rho_\text{cl}$ defined by $P_\text{cl} (\alpha) \geq 0$. 
The action of quadrature squeezing~\cite{Walls1983} $\hat{S}(r) = e^{r(\hat{a}^{\dagger 2} - \hat{a}^2)/2}$ transforms this into a squeezed classical state
\begin{equation}
	\rho_\text{sq,cl} = \int \frac{\D^2\alpha}{\pi}~P_\text{cl}(\alpha) \hat{S}(r) \ket\alpha\bra\alpha \hat{S}^\dagger(r) \, ,
\end{equation}
where $r$ is \emph{loosely} known as the squeezing parameter.  Using the transformation properties of the displacement operator under squeezing, the corresponding normally ordered characteristic function \eqref{eq:scf_genform} is given by
\begin{align}
	\chi^{(N)}_\text{sq,cl}(\lambda) & = e^{|\lambda|^2/2} ~\text{Tr}\args{\rho_\text{sq,cl} \hat{D}(\lambda)} = e^{\left [-\nu^2 |\lambda|^2 + \frac{\mu\nu}{2} (\lambda^2 + \lambda^{*2}) \right ] } 
	\nonumber 
	\\
	&\times \int \frac{\D^2\alpha}{\pi}~P_\text{cl}(\alpha) e^{(\nu\lambda - \mu\lambda^*)\alpha - (\nu\lambda^* - \mu\lambda)\alpha^*},
	\label{eq:cf_sqcl}
\end{align}
where $\mu=\cosh r$, $\nu=\sinh r$ and $\tau=\tanh r$.

It must be noted that here, normal ordering corresponds to $s=1$ in the formulation of $s$-ordered phase space distributions as given in Eq. \eqref{eq:scf_genform}. 
However, in general, one can obtain the normal-ordered form of any operator $F(a,a^\dag)$---$\,\bm{:}\!F(a,a^\dag)\!\bm{:}\,\neq F(a,a^\dag)$, by using an interesting technique that involves integration within operator ordering \cite{Fan2010}.

For physically admissible quantum states, wavefunctions---but not true, in general, for corresponding phase-space distributions---are typically well-behaved functions, often belonging to the Schwartz class, i.e., smooth functions with rapid decay at infinity. 
The Fourier transform acts as an isomorphism on this space, ensuring a consistent mapping between phase-space distributions and their characteristic functions.

The classicality condition follows more directly from the $P$-representation itself.  For a classical state, with $P_\text{cl}(\alpha) \geq 0$, 
\begin{equation}
	\chi_\text{cl}^{(N)}(\lambda) = \int \frac{\D^2\alpha}{\pi}~P_\text{cl}(\alpha) ~e^{\lambda\alpha^* - \lambda^*\alpha} \, ,
\end{equation}
wherefrom it follows immediately that~\cite{Vogel2000}
\begin{equation}
	\abs{\chi_\text{cl}^{(N)}(\lambda)} \leq \int \frac{\D^2\alpha}{\pi}~P_\text{cl}(\alpha) = 1.
\end{equation}

Any violation of this bound therefore provides a clear and operational signature of quantumness~\cite{Vogel2000, Richter2002}. 
As a consequence, for a squeezed classical state \eqref{eq:cf_sqcl}, constructed from an initially classical distribution $P_\text{cl}(\alpha)$, this criterion imposes a lower bound on the squeezing parameter $r$.

The case of a general $P_\text{cl}(\alpha)$ is beyond our present scope, but the Gaussian case is instructive. 
Since, displacements do not alter the Gaussianity profile, we focus only on the thermal states which represent Gaussian mixed states with \emph{classical} $P_\text{cl}(\alpha)$; i.e., $P_\text{th}(\alpha) = \frac{1}{\bar{n}}e^{-\frac{|\alpha|^2}{\bar{n}}}$.  
Substituting into \eqref{eq:cf_sqcl} leads to the characteristic function of the squeezed thermal state as
\begin{align}
	\chi^{(N)}_\text{sq,th}(\lambda) & = e^{-\args{(\mu^2+\nu^2)\bar{n} + \nu^2}|\lambda|^2 + \mu\nu \argp{\bar{n}+\frac{1}{2}} \argp{\lambda^2 + \lambda^{*2}} }.
	\label{eq:cf_sqth}
\end{align}
Now, by changing to real variables as $\lambda = (\lambda_1 + i\lambda_2)/\sqrt{2}$, $\abs{\chi^{(N)}_\text{cl}(\lambda)} \leq 1$ immediately yields
\begin{align}
	\frac{\nu}{\mu} \leq \frac{\bar{n}}{\bar{n} + 1},
	\label{eq:sqth_class_cond}
\end{align}
leading to the well-known condition for the \emph{quantumness} of squeezed thermal state $r > \frac{1}{2}\ln (2\bar{n} + 1)$.  The lesson is that while squeezing can certainly drive a state into the quantum regime, the squeezing strength required to do so depends on how classical---how thermally mixed---the initial state already was.

\noindent
{\bf Quantumness induced by Kerr interaction}---
Consider now  the Kerr nonlinear interaction~\cite{Yurke1986, Kitagawa1986}, for which the corresponding propagator is given as $\hat{\mathcal{K} }(\gamma) = e^{i\gamma \hat{N}(\hat{N}-1)}$, such that $\gamma$ stands for interaction strength and $\hat{N} = \hat{a}^\dagger \hat{a}$ is number operator. 
The action of such interaction on a coherent state is given by
\begin{align}
	\hat{\mathcal{K}}(\gamma) \ket\alpha & = e^{-\frac{|\beta|^2}{2}} \sum_k \frac{\beta^k}{\sqrt{k!}} e^{i\gamma k^2} \ket k,
	\label{eq:kerr_pure}
\end{align}
where $\beta = \alpha e^{-i\gamma}$. 
The nontrivial phase factor makes the state \eqref{eq:kerr_pure} different from a coherent state. 

A simple and elegant example of this kind was first explored by Yurke and Stoler \cite{Yurke1986}, and independently by Kitagawa and Yamamoto in the context of \emph{amplitude squeezing} (we shall discuss later) \cite{Kitagawa1986}. 
In these references, a coherent state under Kerr nonlinear interaction undergoes self-phase modulation and can potentially yield a superposition of coherent states, often referred to as a ``Schr\"odinger cat state.''

This is applied not only to pure states but also to a large class of mixed states, as the corresponding evolution departs from a Fokker-Planck-type equation~\cite{Drummond1980}. 
It was theoretically shown \cite{Jeong2006, Jeong2007} and experimentally demonstrated \cite{Yang2025} that a Kerr nonlinear interaction transfers displaced thermal states to mixed-state versions of Schr\"{o}dinger cat states.
From this perspective, Kerr squeezing \cite{Yurke1986, Kitagawa1986} is distinct from the quadrature squeezing \cite{Walls1983}.

Since the coherent state is the \emph{only possible classical pure state} \cite{Hillery1985}, this result immediately hints at the presumption that every Kerr-evolved pure states are necessarily quantum.
Furthermore, in view of the above mentioned observations on mixed states, it is provocative to extend the idea as \emph{Kerr non-linear evolution guarantees quantumness}.
However, it is not true, in general.

To appreciate this, one needs to take a closer look at the impact of Kerr non-linear evolution, as given in Eq. \eqref{eq:kerr_pure}.
It primary introduces a phase-modulation onto the number states. 
Hence, in absence of any superposition between different number states, the additional state-dependent phase-factors vanish under conjugation, i.e., when we consider the density operator. 
This, trivially leads to the observation that the \emph{Kerr non-linearity fails to introduce quantumness onto the input states, in absence of superposition between number states.}
The vacuum state and the thermal state provide such simple counterexamples.

To conclude, we note that operations such as photon subtraction, quadrature squeezing and amplitude squeezing, which need not produce quantumness alone, do so when combined with others, regardless of the input state---for instance, coherent addition and subtraction. 
Note, however, that \emph{such operations do not always reinforce each other}: in some combinations they compete, reducing the overall quantumness \cite{Bose2017}.

\section{Signature of quantumness}
\label{sec:signature}
Quantumness is formally defined through irregular features of phase-space distributions—negativity or singularity of the $P$ function, or negativity of the Wigner function---a much stricter condition than the former. 
Yet over the past seven decades, this abstract definition has repeatedly surfaced in concrete experimental signatures, without requiring full state tomography. 
Unveiling these quantum fingerprints remains an active and evolving area of research.

\noindent
{\bf Sub-Poissonian statistics---}
The most elementary measurement on an optical signal is  intensity, which in the quantum domain corresponds to the number operator $\hat{N}=\hat{a}^\dagger \hat{a}$. 
Photon-counting statistics therefore provide a simple and powerful probe of quantum character \cite{Luoisell_Book1973}.
A coherent state obeys Poissonian~statistics,
\begin{equation}
	P_{\ket\alpha} (n) = |\langle n\vert \alpha\rangle|^2 = e^{-\lambda} \frac{\lambda^n}{n!},
\end{equation} 
where $\lambda = |\alpha|^2$.

Any classical mixture over coherent states will only \emph{broaden} this distribution histogram, yielding a super-Poissonian distribution. 
The converse—a distribution narrower than Poissonian, termed sub-Poissonian \cite{Mandel1979}---is therefore impossible for any classical state \cite{Luoisell_Book1973} to attain. 
The photon-number state, an energy eigenstate of the harmonic oscillator, is the paradigmatic example.    
More generally, finer criteria can be built from correlations between successive terms of the photon-number distribution, as developed over several decades~\cite{Klyshko1996, Arvind1996, Shchukin2005, Luis2015, Luca2022, Griffet2023, Kovtoniuk2024}.

\noindent
{\bf Photon antibunching---}
A related but distinct signature appears in the temporal correlations between photons. 
Since light is a bosonic field, photons expected to \emph{bunch} (come in pairs). 
This tendency is captured by the second-order coherence function
\begin{equation}
	g^{(2)}(0)  = \frac{\av{\hat{a}^{\dagger 2} \hat{a}^2}}{\av{\hat{a}^\dagger \hat{a}}^2},
	\label{eq:g2}
\end{equation}
which equals exactly one for a coherent state and exceeds one whenever photons preferentially arrive in pairs. 
It is interesting to note that $g^{(2)}(0)$ in Eq. \eqref{eq:g2} is exactly same as the correlation function used in Eq. \eqref{eq:g2tau} for Hanbury Brown--Twiss experiment.
Any classical light is ought to follow the condition $g^{(2)}(0)\geq 1$.

On the other hand, for a photon number state $\ket n$, it is trivial to see that $g^{(2)}(0)<1$. This implies that photons arrive individually rather than in groups, a phenomenon known as antibunching~\cite{Kimble1977} and one of the clearest departures from classical optics \cite{Luoisell_Book1973} (see \cite{Paul1982} for a detailed discussion on antibunching).  Mandel later modified the $g^{(2)}(0)$ and defined another normalized quantity that provides more direct observation of quantumness, known as the Mandel $Q$ parameter~\cite{Mandel1979}
\begin{equation}
	Q = \frac{ \langle \argp{\Delta \hat{n}}^2  \rangle - \av{ \hat{n} } }{ \av{ \hat{n} } } = \av{ \hat{n} }  [g^{(2)}(0) - 1] .
	\label{eq:mandelq}
\end{equation}

For classical states admitting a nonnegative Glauber–Sudarshan P distribution, $g^{(2)}(0)\geq1$ and hence $Q\geq 0$. 
Conversely, $Q<0$, corresponding to sub-Poissonian photon statistics, is a sufficient signature of quantumness. 
However, $Q\geq 0$ does not necessarily imply classicality; the sqeezed vacuum states are such an example of quantum light.

Despite their similar mathematical form, sub-Poissonian statistics and photon antibunching are not equivalent phenomena~\cite{Zou1990}. 
for a single mode light, both have identical form. 
However, the difference appears in the case of multimode light.
The former probes fluctuations in photon number, while the latter probes temporal correlations between detection events. 
In a nutshell, sub-Poissonian statistics: suppressed number fluctuations, i.e., light more precise than any classical source; whereas antibunching: photons avoid arriving together.

\noindent 
{\bf Squeezing---} 
While photon statistics requires the full photon-number distribution, quantum signatures can also be extracted from second-order moments---also known as variance---of field observables alone.
For a coherent state, both the position and momentum quadrature uncertainties (square root of variance) saturate, i.e., $\Delta x = \Delta p = 1/2$.
Moreover, for any quantum state that can be represented by a positive (classical) mixture of coherent states, $\Delta x$ and $\Delta p$ cannot have any value below $1/2$. 
This, in turn, sets the lower bound of uncertainties attainable for any classical state.

However, Heisenberg's relation only constrains the product of conjugate uncertainties, $\Delta x \Delta p \geq 1/4$, leaving the individual uncertainties unconstrained. 
A reduction below 1/2 in one quadrature (squeezing) must therefore be compensated by a corresponding increase in the other (antisqueezing)~\cite{Walls1983}. 
As a consequence, one can simply observe the quantum effect of quadrature squeezing by checking if $\Delta x$ (or $\Delta p$) is less than $1/2$ \cite{Walls1983}.  
One may find a good review on this topic at \cite{Dodonov2002} and their experimental aspects at \cite{Andersen2016}.

While the uncertainty relation is usually realized in terms of the canonical variables like \emph{position}-\emph{momentum}, it can also be expressed in terms of other conjugate variables such as energy-time, number-phase, etc. 
As a consequence, observing uncertainty along any of these canonical variables below the classical limit serves as the quantumness indicator. 
From \eqref{eq:kerr_pure}, it is evident that the Kerr nonlinearity introduces self-phase modulation on the coherent states, e.g., $\alpha \rightarrow \alpha e^{-i\gamma}$, yielding an increase in uncertainty in \emph{phase}. 
It is only natural to think of an decrease in the number operator to compensate for the same.
However, the relation between uncertainties in number and phase is bit more intrigued than that of quadrature variables.
Under a particular construction of operators representing the effective number ($\hat{n_c}$) and phase ($\hat{\Phi}$), in an interferometric setup \cite{Kitagawa1986}, it can be shown that similar to the quadrature variables, there exists another uncertainty relation $\Delta n_c \Delta \Phi \geq 1/2$. 
From the crescent shapes of Kerr-evolved coherent states, it is clear that the state possesses squeezing along the amplitude and thus the name \emph{amplitude squeezing}. 
From this perspective, a photon number state is an example of extreme amplitude squeezing such that \emph{the crescent shape turns into a perfect circle}. 

\noindent
{\bf Higher-order moments---}
Some states exhibit neither sub-Poissonian statistics nor ordinary squeezing \cite{Tara1992}, making their quantum character harder to detect at the level of second-order quantities. 
This motivates extending both the uncertainty relation and the sub-Poissonian criterion to higher-order moments, exposing quantum features invisible at lower orders.  	

In the context of squeezing, one can think of two different types of such generalizations.  The first one~\cite{Hillery1987}, considers $m^\text{th}$-order quadrature variables like $\hat{X}_m = (\hat{a}^m + \hat{a}^{\dagger m})/\sqrt{2}$ and $\hat{P}_m = - i (\hat{a}^m - \hat{a}^{\dagger m})/\sqrt{2}$ such that $[\hat{X}_m, \hat{P}_m ] = \frac{i}{2} \hat{C}_m$, amplitude-powered squeezing criteria  leads to 
\begin{equation}
	\langle\argp{ \Delta X_m }^2\rangle < \frac{1}{4} \mid \langle \hat{C}_m \rangle \mid.
\end{equation}
	
On the other hand by considering the $m^\text{th}$ power of the $\Delta \hat{x}$ ($\Delta \hat{p}$) one can also define the higher-order squeezing criteria as \cite{Hong1985}
\begin{equation}
	 \langle \argp{\Delta x}^{2n} \rangle  < \langle \argp{\Delta x}^{2n} \rangle_{\ket \alpha} = \frac{(2n)!}{2^{2n} n!}\frac{1}{2^n}.
\end{equation}

Similar to squeezing, one can also define the generalization of the sub-Poissonian character and antibunching \cite{Verma2010}.
However, in general, one can obtain such quantum criteria in terms of any operator \cite{Richter2002} or the moments of the phase-space distributions~\cite{Bohmann2020}.
While all these manifestations of quantumness depend crucially on the detector efficiency, one may further think of a detector-agnostic signature of quantumness~\cite{Bohmann2019}.

\section{Measures of quantumness}
\label{sec:measure}

\subsection{Hilbert-space based approaches}
Since quantum $P$~functions can neither be measured experimentally nor computed in most cases, several indirect methods, grounded in Hilbert space, have been proposed to successfully discern the quantumness. 

\noindent
\textbf{Distance from classical states---}
A direct consequence of the Hahn-Banach theorem \cite{Book_ReedSimon2013} is the geometrical existence of a hyperplane that separates the convex set from any element outside the set. 
Since all classical states form a convex set, this naturally suggests using the minimal distance between any quantum state and the set of classical states as a quantumness measure, paralleling how entanglement witnesses arise from the separation between separable and entangled states~\cite{Terhal2000, Horodecki2009}.
Among the \emph{bona fide} distances, the most popular ones are the Hilbert-Schmidt distance~\cite{Hillery1987a, Dodonov2000, Dodonov2003, Nair2017}, the Bures metric~\cite {Marian2002}, and the trace norm \cite{Park2021}; a comparative analysis of these measures can be found in \cite{Sperling2015}. 
It may be noted that, for the mixed Gaussian states, the Bures-metric-based quantumness is equivalent to the \emph{quantum depth} \cite{Marian2004}, which we shall discuss in the next subsection. 
However, all such distance functions require convex optimization over the set of all classical states, which demands heavy computational resources. 

\noindent 
\textbf{Other proposals---}
Given that coherent states are the only classical pure states \cite{Hillery1985} and that any nontrivial superposition of states exhibits quantum character, it is natural to encode the quantumness of a pure state in terms of the number of distinct coherent-state superpositions~\cite{Vogel2014}. 
This also allows one to appreciate the similarity between quantumness and entanglement by mapping this distinct coherent superposition to the Schmidt rank. 

Alternatively, generalized entropic measures such as Tsallis entropy~\cite{Sadeghi2012} have been proposed to distinguish optical states by their quantum statistics, highlighting how the gap between quantum and classical-like entropies can reveal quantumness~\cite{Zhang2021}. 
Moreover, in the context of Hilbert's $17^\text{th}$ problem, people have encoded the quantumness as the minimum order of a squared polynomial needed to detect the non-regularity of the Sudarshan-Glauber $P$-function~\cite{Korbicz2005, Ohst2025}. 
Related formulations cast the problem in terms of a state's sensitivity to normal ordering of the operators~\cite{Gehrke2012, DeBievre2019, Bohmann2020a}. 
Distance-based measures related to the notion of operator-ordering sensitivity~\cite{DeBievre2019} in relation to the R\'enyi entropy were also~proposed.

\subsection{Phase-space based approaches}

\noindent
\textbf{Gaussian smoothing of $P$-function---}
It can be easily appreciated from \eqref{eq:psd_interrelation} that while the $P$-function might become highly singular to be expressed in terms of ordinary functions, a proper Gaussian transformation can render it both measurable and strictly positive. 
Building on the result by Iagolnitzer~\cite{Iagolnitzer1969}, which states that any Gaussian smoothing of the Wigner function---provided the Gaussian is at least as wide as the minimum uncertainty wave packet---yields a positive semidefinite function, Lee~\cite{Lee1991} introduced a novel approach to quantify the quantumness of a quantum state. 
He proposed that the \emph{quantum depth} of a state be defined as the minimum amount of Gaussian smoothing required on its $P$-function to render it positive semidefinite. 
Mathematically, the depth is the smallest value of $\omega$ ($0 < \omega \leq 1$) in
\begin{equation}
	R(z,\omega) = \frac{1}{\omega} \int \dfrac{\D^2\alpha}{\pi}~e^{-\frac{|z-\alpha|^2}{\omega}} ~P(\alpha),
	\label{eq:depth}
\end{equation}
for which $R(z,\omega) \geq 0$.  
In line with \eqref{eq:psd_interrelation}, it is evident that for all Gaussian states, pure or mixed, depth maximizes at $1/2$ \cite{Lee1991} as it yields the corresponding Wigner function; i.e., $R(z,\omega)\vert_{\omega=1/2}^\text{Gauss} = W^\text{Gauss}(z,z^*)$, which is positive semidefinite. 
One may also appreciate the result: $\omega_{\max} = 1/2$ for Gaussian states, from its connection with the variance matrix \cite{Marian2002, Marian2004} as
\begin{equation}
    \omega = \max \args{0,\frac{1}{2} - \lambda_{\min}},
\end{equation}
where $\lambda_{\min}$ is the minimum eigenvalue of the variance matrix.
This could be seen as a natural extension of Hudson's theorem~\cite{Hudson1974}--that says Wigner positivity for pure states is possible only for Gaussian cases---to the mixed states. 

However, the proposal lacked generality. 
It can be shown that for all pure non-Gaussian states the minimal width is unity; i.e., $\omega=1$~\cite{Lutkenhaus1995} that yield Husimi-$Q$ function. 
As a consequence, it fails to differentiate between different non-Gaussian pure states in terms of their degree of quantumness.  
Nonetheless, it triggered the query of finding suitable smoothing for $P$-function to encode the quantumness including generalization of depth to multimode cases~\cite{Sabapathy2016} as well as to quantum processes~\cite{Keshari2013}. 
An extensive effort in achieving \emph{a general resource theoretic and mathematically consistent} way of regularization of $P$ came from various authors, as we discuss next.

\noindent
\textbf{Filtering $P$-function---}
An alternative to Gaussian smoothing of the $P$-function directly is to filter the characteristic function instead. 
As shown in Fig. \ref{fig:WignerWeyl} and mathematically expressed in~\eqref{eq:psd_opfourier}, it is evident that the  $P$ function is obtained through a Fourier transform of the normal-ordered characteristic function $\chi^{(N)}(\lambda)$. 
However, $\chi^{(N)}(\lambda)$ is nothing but the Weyl characteristic function, $\chi(\lambda) = \text{Tr}\args{\rho\,\hat{D}(\lambda)}$ multiplied by a filter $e^{\frac{|\lambda|^2}{2}}$~\eqref{eq:spsd_genform}. 
This motivates introducing a family of \emph{filter functions}, parametrized by $\omega$, applied before the Fourier transform: $P_\omega(\alpha) := \mathcal{F}\args{\chi^N(\lambda) \Omega_\omega(\lambda)}$ where $\Omega_\omega(\lambda,\lambda^*)$ is the additional filter function. 
Since the goal is to eliminate singularities in the \emph{deformed} $P$-function, the filter cannot be chosen arbitrarily.  
Building on Klauder's early work~\cite{Klauder1966}, Kiesel and Vogel~\cite{Kiesel2010} proposed  three governing criteria:
\begin{itemize}
	\item Universality---the filter $\Omega_\omega(\lambda)$ must apply uniformly to all states.
	
	\item Non-negativity---the Fourier transform of $\Omega_\omega(\lambda,\lambda^*)$ must be non-negative, since otherwise the filter could introduce spurious negativity into the $P$function of a classical state, violating universality.
	In other words, it might yield a negative $P$-function for classical states, which is not possible.
	
	\item Completeness---the filter $\omega$ must act as a scaling factor, such that with infinite width, the deformed $P$-function must return the original $P$-function, i.e., $\lim_{\omega\rightarrow \infty} P_\omega(\alpha) = P(\alpha)$. 
\end{itemize}

Choosing $\Omega_\omega(\lambda)$ to be Gaussian immediately reveals the non-regularity/singularity in the $P$-function in terms of negativity~\cite{Kiesel2010} as well as provides a sufficient criterion to detect quantumness~\cite{Kuhn2018}.
Moreover, Tan \emph{et. al.}~\cite{Tan2020} revealed a remarkable property of such Gaussian filter functions in terms of their capability, under a more restrictive criterion, to \emph{map singular $P$-function to $P$-negativity} that encodes the coherence of the state~\cite{Tan2017}.
These characteristic-function-based approaches also capture superposition~\cite{Ryl2017}, and an elegant Wick-ordered description of the characteristic function for quantifying quantumness appears in~\cite{Li2022}.

\noindent
\textbf{Quantum channel perspective}---
Consider a state defined by $P(\alpha)$ and its normal-ordered characteristic function  $\chi^{(N)}(\lambda) $. 
Now under a filter operation described by $\Omega_\omega(\lambda) = e^{-\omega|\lambda|^2}$, the deformed $P$-function becomes
\begin{align}
	P_\omega(z) = \frac{1}{\omega} \int \dfrac{\D^2\alpha}{\pi} P(\alpha) e^{-\frac{|z-\alpha|^2}{\omega}}.
	\label{eq:def_P}
\end{align}

Evidently, \eqref{eq:def_P} represents the quantum depth~\eqref{eq:depth}, where $\lim_{\omega\rightarrow 0}$ renders $\Omega_\omega(\lambda)$ to be a delta function that recovers the original $P$-function. 
Since $(1/\omega) e^{-\frac{|z|^2}{\omega}}$ is the Fourier transform of $e^{-\omega|\lambda|^2}$, one can immediately generalize the regularization of $P$-function by replacing $\frac{1}{\omega} e^{-\frac{|z|^2}{\omega}}$ with the Fourier transform of the non-Gaussian filter $\Omega_\omega^\text{nG}(\lambda)$~\cite{Kuhn2018}. 
This implies that the \emph{smoothing at the $P$-function is equivalent to applying a filter at the characteristic function level}.

Now consider that  the state $\rho$  undergoes a general (not necessarily Gaussian) filter operation  $\Omega_\omega(\lambda)$  satisfying the above criteria~\cite{Kuhn2018}, with Fourier transform $\mathcal{F}\args{ \Omega_\omega(\lambda) } = \Omega_\omega^\mathcal{F}(z)$.  Changing variables $\alpha\rightarrow \alpha+z$ in~\eqref{eq:def_P}, one can show~\cite{Kuhn2018}
\begin{align}
	\Omega_\omega(\lambda) :\rho~\mapsto \int \frac{\D^2z}{\pi}~\Omega_\omega^\mathcal{F}(z) \hat{D}(-z) ~\rho~ \hat{D}^\dagger(-z),
	\label{eq:filter_stocdis}
\end{align}
showing that filter operations amount to a linear map inducing stochastic displacements, with probabilities drawn from the Fourier transform of the filter function. 
Together, \eqref{eq:def_P} and \eqref{eq:filter_stocdis} establish the equivalence of three pictures: phase-space smoothing, characteristic-function filtering, and linear-optical maps inducing stochastic displacement.

This raises a further question: which maps regularize the $P$-function? For Gaussian maps, it was recently shown that the positivity of $P$-function after regularization can be achieved~\cite{Ivan2017, Zartab2023} by composing \emph{quantum limited amplifier} and \emph{quantum limited attenuator} channels that transform the input $P$-function into Husimi $Q$-function~\cite{Linowski2024}.

Closely related to this quantum channel perspective, researchers have introduced the notion of the \emph{minimal additive noise channel} required to render the $P$-function positive---a quantification of quantumness known as robustness~\cite{Regula2021}.  For any state $\rho$, it is defined as
\begin{equation}
	\mathcal{R}(\rho):= \inf_{\sigma \in \Omega_\text{cl}} \argc{r|r\geq 0 ~:~ \frac{\rho + r\sigma}{1+r}\in \Omega_\text{cl}}
	\label{eq:robustness}
\end{equation}
where $\Omega_\text{cl}$ is the set of classical states. 
While robustness, in general, requires convex optimization over all such channels, some partial results can be obtained for Gaussian states and pure states. 
Nonetheless, in view of the result that any single-mode Gaussian state is a \emph{displaced squeezed thermal state}~\cite{Chaturvedi1989}, one can also envisage the robustness as a generalization of quantum depth~\cite{Lee1991}.

\noindent
\textbf{Exploiting Wigner function}---
Among all possible $s$-ordered distributions~\cite{Cohen1966}, only Wigner function reproduces the correct marginals~\cite{Takahashi1986, Takahashi1989}; i.e.,
\begin{equation}
	\int \D p \, W(x,p) = |\psi(x)|^2, \quad \int \D x \, W(x,p) = |\phi(p)|^2 \, ,
\end{equation}
where $\psi(x)$ and $\phi(p)$ are the position and momentum wavefunctions. 
This property has made the Wigner function central to characterizing quantum behavior.

While the Wigner function for a coherent state is just a Gaussian with equal width along both quadratures, any classical state---built from a positive distribution over coherent states---also has a positive Wigner function. 
As with the $P$-function, any departure from positivity indicates quantum character, leading to a straightforward quantification in terms of the total negative volume of the Wigner function~\cite{Kenfack2004}.  
A related approach links the Wigner-Yanase skew information directly to the quantum character~\cite{Luo2019, Tang2025}. 
The one-to-one mapping between density operators and Wigner functions also allows coherent superpositions to be read off as interferometric oscillations in the Wigner function~\cite{Sadeghi2010, Lee2011}. 
Finally, optical homodyne tomography, which reconstructs the Wigner function experimentally, enables quantumness measures based on coarse-graining~\cite{Roh2023} and on neural-network-based characterization~\cite{Gebhart2021}.

\noindent
\textbf{Husimi-based approaches---}
Return to the definition of a classical state in~\eqref{eq:def_classical}, where $f_\text{dis}(\alpha) \geq 0$.  Written this way, every classical state can be sampled---and hence simulated---as a mixture of coherent states.  Two properties that make such a sampling description work: \emph{positivity} and \emph{regularity}.  General quantum states possess neither in the diagonal representation and so admit no classical sampling description at all.  Since every state of light carries both quantum and classical traits, it is only natural to ask whether some \emph{sampling distribution} might sit at the interface between the two.

The Husimi $Q$-function answers this.  Among all the Gaussian-smoothed quasidistributions, it is the least smoothed that remains non-negative for every state, and so the closest positive counterpart to the diagonal representation.  Its appeal is not merely mathematical convenience: the $Q$-function is precisely what a simultaneous measurement of position and momentum---heterodyne detection---returns ~\cite{Arthurs1965, QDist_Arthurs_PRL1988}.  This grants it an operational interpretation as the classical-like sampling distribution of a quantum state.

For continuous variables, another interesting property that is closely tied to quantumness is non-Gaussianity. 
Any computation confined to Gaussian states and Gaussian operations can be simulated efficiently on a classical computer, which makes departure from Gaussianity the indispensable resource for advantage in this setting. 
The stellar representation~\cite{Chabaud2020} quantifies that resource operationally through the stellar rank: the number of zeros of the Husimi $Q$-function, equivalently the minimal number of photon additions needed to build the state from a Gaussian seed. 
The rank is invariant under Gaussian unitaries and induces a strict hierarchy over single-mode states. 
Its recent extension, the approximate stellar rank, accommodates realistic near-target state conversions and yields explicit no-go results for Gaussian distillation protocols of direct relevance to bosonic error correction~\cite{Hahn2024}.

The same idea---reading a state through the zeros of its Husimi function---carries over to finite dimensions. 
There it becomes the Majorana stellar representation~\cite{Majorana1932,Bacry2004,Devi2012,Bengtsson2017}, which maps any spin-$S$ state to a constellation of $2S$ points on the Bloch sphere and thereby lends a geometric reading to symmetry, structure, and entanglement content. 
Constellations spread as widely as possible sit farthest from every spin coherent state and are, in this precise geometric sense, the \emph{most quantum}; constellations clustered in a small region of the sphere are the least~\cite{Sanchez2026, Kam2026}. 
Seen through quantum polarimetry, these most quantum states are the ones dominated by higher-order Stokes operators beyond the dipole expansion, and they have been shown to sharpen metrological precision appreciably in parameter estimation~\cite{Bouchard2017, Goldberg2021, Goldberg2022, Goldberg2022a}. 
Such notions tighten the link between discrete-variable quantum information and the geometric picture offered by phase-space representations.

The $Q$-function also underwrites a notion of classical entropy \cite{Wehrl1978}. 
Its departure from the quantum case is striking: unlike the von Neumann entropy, the Wehrl entropy never vanishes, not even for pure states~\cite{Lieb1978}. 
Its connections to the standard quantum entropy~\cite{Wehrl1979, Uhlmann1980} and to the sampling picture \cite{Buzek1995, Buzek1995a} subsequently gave rise to an entropic formulation of quantumness \cite{Bose2019}, later expressed as an excess entropy \cite{Zhang2021} and refined into a sharper characterization of the extremal states~\cite{Goldberg2020}.
			
Besides its experimental feasibility, $Q$-function provides an alternate and efficient~\cite{Muller2016} platform to analyze various physical phenomena, such as understanding the complexity of state~\cite{Sugita2002}, phase transition \cite{Wang2021}, quantum dynamics \cite{Holmes2023}, as well as entanglement-detection~\cite{Floerchinger2022, Garttner2023}. 

\subsection{Resource-theoretic approaches}
\noindent
\textbf{Quantumness as a resource in entanglement generation}---Take a state with a diagonal representation~\eqref{eq:SGP_function}.  A balanced beam splitter divides an incident coherent state evenly between its two output ports, the second input being vacuum, so that $\ket\alpha \xrightarrow{BS} \ket{\alpha/\sqrt{2}} \ket{\alpha/\sqrt{2}}$.  Acting on $\rho$, it therefore produces a diagonal representation that is manifestly separable whenever $P(\alpha)\geq 0$.   Inseparability---the impossibility of writing the joint state as a statistical mixture of product states---is thus inherited directly from the non-regularity of the input $P$-function.  In a passive linear-optical network, \emph{quantumness and entanglement are interconvertible}, as illustrated through examples in~\cite{Kim2002, Wang2002}.

The necessity of quantumness for entanglement follows from a short quantitative argument. 
Applying the negative partial transposition (NPT) criterion~\cite{Peres1996} to $\rho_\text{out}$, and noting that partial transposition conjugates the coherent amplitude of the second mode, one finds
\begin{align}
	&\mathcal{N}\argp{\rho_\text{out}} = \left\Vert \rho_\text{out}^\text{P.T.} \right\Vert_1
	\nonumber 
	\\
	&= \left \Vert \int \frac{\D^2\alpha}{\pi} P(\alpha)  \,  |\alpha/\sqrt{2}\rangle \langle \alpha/\sqrt{2} |  \otimes |\alpha/\sqrt{2}\rangle \langle \alpha/\sqrt{2} |  \right\Vert_1
	\nonumber 
	\\
	&\leq  \int \frac{\D^2\alpha}{\pi}  |P(\alpha)| \,  \left\Vert | \alpha/\sqrt{2}\rangle \langle \alpha/\sqrt{2} |  \otimes |\alpha/\sqrt{2}\rangle \langle \alpha/\sqrt{2} |  \right\Vert_1    	
	\nonumber 
	\\
	&=  \int \frac{\D^2\alpha}{\pi}  \abs{P(\alpha)} \, ,
	\label{eq:negativity_bsout}
\end{align}
where $||.||_1$ stands for trace norm and ``P.T.'' for partial transposition such that $\mathcal{N}>1$ indicates inseparability. 
Equation~\eqref{eq:negativity_bsout} makes the implication immediate: $\mathcal{N}\argp{\rho_\text{out}}>1$ forces $1/\pi \, \int \D^2\alpha \,  |P(\alpha)| > 1$, which no nonnegative $P$-function can satisfy. 
The input must therefore have a $P$-function that is somewhere negative, or singular enough that the integral diverges. 
This observation led to the first proposal to quantify input quantumness by the entanglement it generates, measured by an additive entanglement monotone (although not convex), the logarithmic negativity~\cite{Asboth2005}. 
However, whether quantumness is also \emph{sufficient} for entanglement remained open.

That question was settled later~\cite{Ivan2011}: by recasting it as a Stieltjes moment problem, the authors showed that input quantumness is both necessary and sufficient for NPT at the output of a linear-optical setup. 
The result opened the way to quantifying quantumness, understood as quantum superposition, through the entanglement it produces---by the Schmidt rank for pure states~\cite{Vogel2014}, and subsequently as a means of ordering states~\cite{Sperling2015}.  
Entanglement monotones applicable to mixed states have been studied~\cite {Miranowicz2015, Kadlec2024}, and comparisons among these measures are available in~\cite{Li2010}. 
The duality between optical quantumness and inseparability~\cite{ Killoran2016} extends further still to steering and Bell nonlocality \cite{Arvind1999, Park2015, Kadlec2024_2}, so that quantumness may be quantified indirectly through these stronger manifestations of quantum correlations.

\noindent
\textbf{Quantumness as a resource in metrology}---Let the same state~\eqref{eq:SGP_function} undergo a unitary transformation that imprints a parameter $\theta$, 
\begin{equation}
	\rho \rightarrow \rho_\theta = U_\theta \rho U_\theta^\dagger =  \int \frac{\D^2\alpha}{\pi} \, P(\alpha) U_\theta \ket\alpha\bra\alpha U_\theta^\dagger,
\end{equation}
where $U_\theta = e^{i\theta\hat{G}_\theta}$ and $\hat{G}$ is the generator of the transformation. 
It is well-known that any unbiased estimator of $\theta$ is bounded from below by the quantum Cram\'er-Rao bound (QCRB)~\cite{Helstrom1969} as
\begin{equation}
	\text{Var}(\theta) \geq \frac{1}{M F_q(\rho_\theta)},
\end{equation}
where Var stands for variance (square of uncertainty), $M$ is the number of trials and $F_q(\rho_\theta)$ is the quantum Fisher information~(QFI).

For a pure state the QFI reduces to $F_q\big\vert_{\rho = \ket\psi\bra\psi} = 4\,\text{Var}_\rho (\hat{G})$.
Because the QFI is convex, a classical state with $P(\alpha)\geq 0$ can do no better than the average over its coherent-state constituents,
\begin{equation}
	F_q(\rho_\theta)\big\vert_{\rho_\text{cl}} \leq 4 \int\frac{\D^2\alpha}{\pi} P(\alpha)\, \text{Var}_{\ket\alpha} (\hat{G}) ,
\end{equation}
so no mixture of coherent states can sharpen the measurement beyond what a coherent state already achieves. 
Quantumness is therefore necessary for enhanced sensing~\cite{Giovannetti2004}. 
This furnishes an operational route~\cite{Rivas2010, Yadin2018} to gauging the quantumness of a state by how far it raises the QFI---equivalently, how far it lowers the QCRB below the standard quantum limit, while still respecting the Heisenberg limit that constrains any quantum system~\cite{Ou1997}.

Building on this, sharper estimates and a convex resource theory of quantumness in terms of the QFI have been developed~\cite{Hyukjoon2019, Ge2020}, along with schemes that quantify quantumness and entanglement simultaneously for phase estimation~\cite{Wei2022}.

\begin{figure*}
    \begin{center}
    \includegraphics[width=2.1\columnwidth]{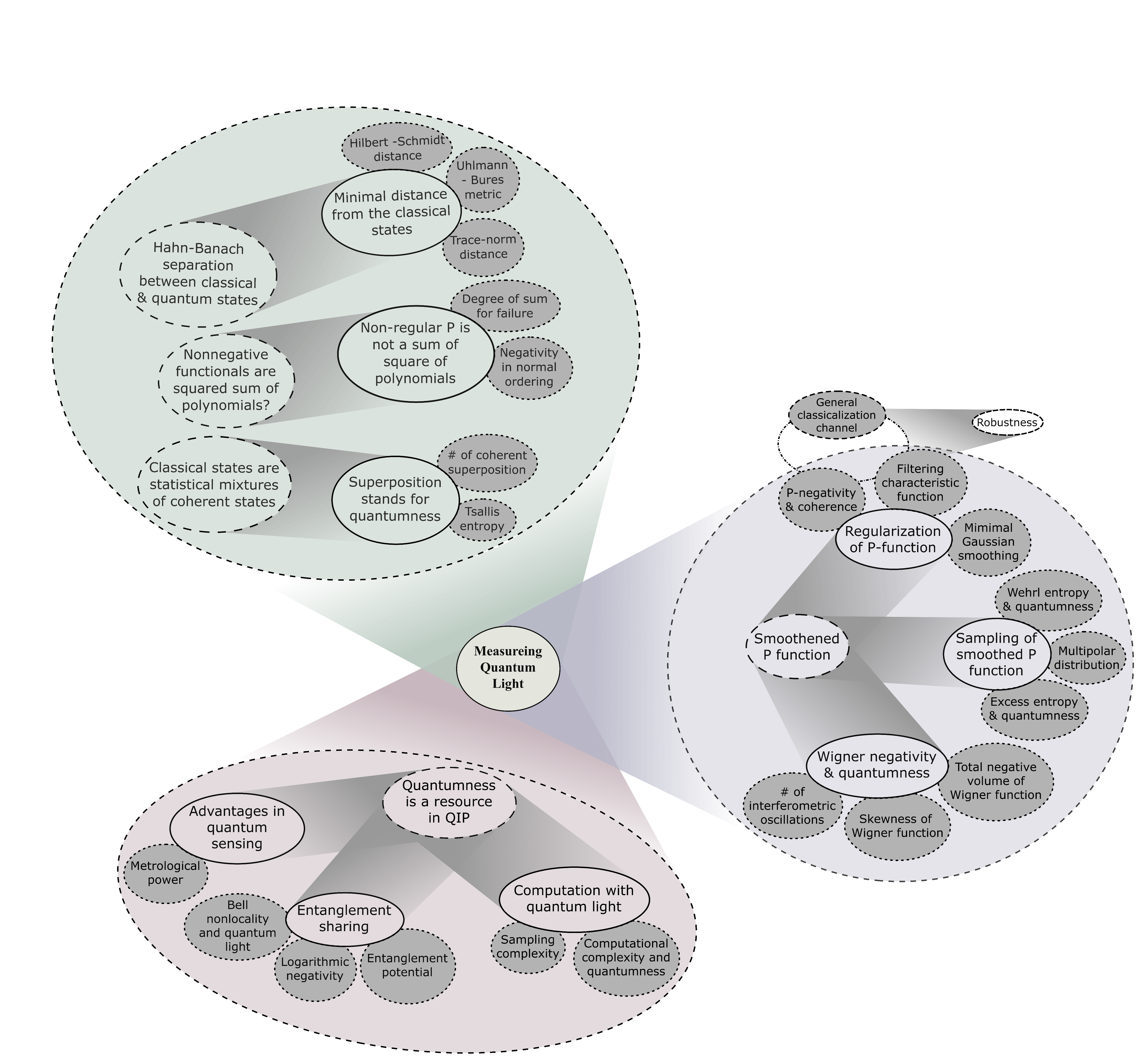}
    \caption{Summary of various approaches in quantifying the optical quantumness.
    Different measures have been divided into three major groups based on their basic character and operational perspectives, such as {\bf (a)} Hilbert-space based approaches, {\bf (b)} phase-space based approaches, and {\bf (c)} resource-theoretic approaches.
    Underlying principles have been encircled with dashed lines, while their manifestation is represented in circles with solid lines.
    Satellite dotted-gray circles stand for the specific proposal.}
    \label{fig:NcMeasure}
    \end{center}
\end{figure*}

Check Fig. \ref{fig:NcMeasure} for a comprehensive idea about various measures of optical quantumness.

\noindent
\textbf{Quantumness and simulation complexity}---
While classical states with positive $P$ distributions $P(\alpha) \geq 0$ can be efficiently simulated, a departure from the positivity may lead to computational difficulties in classical simulation \cite{Rahimi2016}. 
We note that optical nonclassicality alone does not guarantee classical computational intractability. 
For example, Gaussian states, although they are squeezed, can be classically simulated given Gaussian operations and measurements \cite{Mari2012}. 
Efficient-simulation criteria depend jointly on the input states, transformations, and measurements. 
Efficient classical simulability depends jointly on the input states, transformations, and measurements \cite{Rahimi2016}.

Moreover, the efficiency of boson sampling tracks the quantumness of the light employed \cite{Renema2018}, while its complexity can be tuned through a generalized concurrence \cite{Chin2018}. 
For an introduction from the computational side, see \cite{Aaronson2011}.

\subsection{Macroscopic quantumness}

A related but distinct question concerns the physical scale of quantumness, rather than quantumness alone. It is an intriguing question related to Schr\"{o}dinger's celebrated illustration of a genuinely macroscopic quantum superposition \cite{Schrodinger1935}. Such a measure should quantify both the degree of quantum coherence and the size distinctness between the component states.  A large mean photon number is thus insufficient: a bright coherent state remains classical, and an incoherent mixture of widely separated coherent states lacks
the coherence of their superposition.

An early attempt to characterize macroscopic quantumness was Leggett's notion of disconnectivity \cite{Leggett1980}. 
For bosonic fields, a phase-space measure was introduced that combines the scale and strength of interference structures and is related to the rate of purity change under photon loss \cite{Lee2011}. 
Complementary approaches characterize effective size through quantum Fisher information for suitably normalized collective observables in many-body systems \cite{FroewisDur2012}, or assess the distinguishability of superposed photonic components under coarse-grained photon counting \cite{Sekatski2014}. 
Resource-theoretic criteria for macroscopic coherence have also been proposed \cite{YadinVedral2016}. 
A measure that satisfies these criteria, applicable to both spin and bosonic systems, was suggested based on coarse-grained measurement disturbance, while highlighting the need for an appropriate measurement scale to distinguish microscopic from macroscopic coherence \cite{Kwon2017}.

These approaches need not rank states identically, reflecting different choices of physical observables and operational criteria. 
Macroscopic quantumness should thus be distinguished from $P$-quantumness and Wigner negativity, while providing a complementary perspective on optical cat-like states and their sensitivity to decoherence. 
Comparative discussions of these measures and their implementations can be found in Refs.~\cite{JeongKangKwon2015, Froewis2018}.

\section{Applications of quantum light -- practical information technologies}
\label{sec:application}

\subsection{Applied and foundational quantum optics}
Optical states with a non-regular Sudarshan--Glauber $P$ function, together with the operations that induce quantumness, underpin a wide range of applications in quantum optics and serve equally to test the foundational principles of quantum theory. Squeezed light is the outstanding example: proven optimal for the task \cite{Dobrza2013}, it is now used extensively in laser interferometry \cite{Zuo2020, Genovese2021, Podoshvedov2024}, and further gains in quantum interferometry have been proposed by supplementing it with photon addition or subtraction \cite{Tan2014, Birrittella2014, Ouyang2016, Wang2019, Fassakhova2025}.

The same operations prove critical in quantum illumination~\cite{Zhang2014, Sanz2017, Fan2018, Lee2021, Zhang2021a, Zhang2024, Zhong2025}, quantum imaging~\cite{Lugiato2002, Moreau2019, Hugo2024}, and ghost imaging~\cite{Liu2021}. Photon subtraction, in particular, opens a window onto sub-Planck-scale phase-space structures~\cite{Akhtar2023} and has been used to verify the fundamental commutation relation of quantum theory~\cite{Parigi2007}.

\subsection{Quantum communication}

\noindent
\textbf{Entanglement sharing in linear network}---Quantumness is a prerequisite for generating entanglement and the passive linear beam splitter is the instrument that converts it: single-mode quantumness in, two-mode entanglement out~\cite{Hofmann2007, Afek2010, Croal2015}.  The same device also serves as a tool for analysing majorization problems ~\cite{Gagatsos2013}. 	 The correspondence is not one-to-one, however.  For single-mode states, increasing the quantumness does not always increase the entanglement, whereas Gaussian states obey a conservation relation between the two \cite{Ge2015}.
	
Quantumness-inducing operations have likewise proved valuable for improving the entanglement shared between two parties.
Photon addition and subtraction have been studied individually to this end~\cite{Kitagawa2006, Zhang2010, Lee2011a, Lee2013, Zhang2013, Bartley2013, Benlloch2016, Walschaers2017, Meng2020, Song2023, Bashmakova2023, Zhang2025}, as have various combinations of the two \cite{Wu2015, Yang2016, Yang2019, Mojaveri2019}.

Two general findings are worth noting: the entanglement gain from a single photon subtraction cannot exceed $\log 2$ \cite{Zhang2022}, and subtraction consistently outperforms addition \cite{Bartley2015, Das2016}.

\noindent
\textbf{Quantum teleportation}---
Quantum states and operations matter not only for distributing entanglement between distant parties, but also for what is subsequently done with it. 
For example, quantum teleportation can be efficiently performed using cat states \cite{vanEnk2001, Jeong2001}. 
Alongside squeezing, photon addition and subtraction play a central role in quantum teleportation~\cite{Opatrny2000, Olivares2003, Dellanno2007, Yang2009, Dhar2015, Wang2015, Bose2021, He2025, Arora2025}, tele-amplification~\cite{Neergaard2013, Fiurasek2022, Jie2023, Fiurasek2024}, and state amplification~\cite{Jeon2025}. 
While the quadrature squeezing, besides entanglement, was shown to be necessary for teleportation~\cite{Bose2021a}, later work indicates that, in its absence, other quantum features can take over as the resource for certain non-Gaussian states~\cite{Sharma2025}.
A relevant review is available in~\cite{Shi2023}.

\noindent 
\textbf{Bell-nonlocality and randomness generation}---
Beyond the states fed into a passive linear device, quantumness-inducing operations also help to generate and strengthen nonlocal correlations---which can be much stronger than entanglement---between distant parties. 
Bell nonlocality extends to infinite-dimensional optical states through a variety of settings: parity measurements~\cite{Banaszek1999, Jeong2003, Lee2009a, Abouraddy2007, Yarnall2007}, pseudo-spin observables~\cite{Son2004, Chen2005, Ferraro2005, Singh2025}, homodyne detection~\cite{Paternostro2009, He2010, Kiukas2010, Torlai2013, Chen2014, Park2015, Schneeloch2016, Thearle2018, Freire2019}, the covariance matrix for Gaussian states~\cite{Cavalcanti2007}, angular variables~\cite{Borges2012}, and photon discrimination~\cite{Lezama2023}. 
It has also been analyzed in the hybrid setting involving atom-light entanglement \cite{Brask2012, Hyukjoon2013, Moradi2024, Bera2025, Peizhe2026}.
Finer aspects of continuous-variable Bell nonlocality are treated in~\cite{Weiss2016, Finot2017, Barbosa2022, Carlos2026}. 
These developments have, in turn, enabled practical heterodyne-based semi-device-independent randomness generation~\cite{Marangon2017, Avesani2018, Xu2019, Huang2020, Avesani2021, Tebyanian2021, Li2023, Zhang2025a} and the exploration of generalized probabilistic theories~\cite{Ketterer2018}.

\noindent
\textbf{Quantum cryptography}---
Secure communication between distant parties began with the explicit use of squeezed states~\cite{Hillery2000, Cerf2001, Gottesman2001}, but it later emerged that quantumness, in fact, \emph{is not a prerequisite for security}~\cite{Grosshans2002, Grosshans2003, Usenko2010, Weedbrook2012}. 
It nonetheless confers substantial practical advantages, among them the most important ones are higher key rates and greater resilience to noise~\cite{Patron2009, Usenko2011, Usenko2018, Oruganti2025, Nguyen2025}. 
The benefits of correlated squeezed light~\cite{Ralph1999, Ralph2000, Reid2000, Pereira2000, Silberhorn2002, Madsen2012} and of non-Gaussian modulations~\cite{Leverrier2011} have sustained the pursuit of cryptography with quantum states \cite{Chandan2019, Jaskaran2021}. 
To be precise, in the current pursuit of surpassing the repeater-free communication bound \cite{Plob2017}, quantum repeaters \cite{Fabian2018, Kaushik2020, Fukui2021, Filip2021, Wu2022, Dias2022} and their memoryless modulations \cite{Peizhe2023, Peizhe2024} are of prime interest.
For detailed reviews on the field, see~\cite{Zhang2024a, Motaharifar2025}.

\subsection{Quantum computation}

Besides enhancing the simulability beyond the classical limit, quantum states and operations are also useful in various other facets of practical implementation of quantum computation \cite{Mahmud2025}. 
Fault-tolerant quantum computing with error correction \cite{Knill1997, vanLoock2010} may be performed using the binomial codes \cite{Michael2016, Hu2019}, cat codes \cite{Cochrane1999, Jeong2002, Ralph2003, Ourjoumtsev2006, Ourjoumtsev2007, Lund2008, Mirrahimi2014, Bergmann2016, Hastrup2020, Darmawan2021, Su2022, Hastrup2022, Chamberland2022}, squeezed-cat codes \cite{Gao2018, Teh2020, Schlegel2022, Xu2023, Hillmann2023, Provaznik2025}, GKP states \cite{Gotteman2001, Tzitrin2020, Royer2020, Grimsmo2021, Marek2024, Rojkov2024}, hybrid states ~\cite{LeeJeong2013, Jeong2014, Morin2014, Omkar2020, Omkar2021, Jaehak2024}, rotation-symmetric-bosonic-codes (RSBCs) \cite{Grimsmo2020} as well as number-phase encoding \cite{Hu2025}. 
One may find a detailed review on the topic in ~\cite{Jaehak2025}. 
Furthermore, quantum error mitigation/suppression with cat states
has recently been explored 
\cite{Shringarpure2024, Teo2025, Endo2025} .

Apart from the explicit use of systems possessing strong ``quantumness'' as physical computation and error-correction resources, this notion is also intimately tied to the progress of the field of quantum~computation. 
Owing to a wealth of existing contributions in this field related to the notion of~quantumness, we shall list and summarize only key topics of discussion about their connection:

\noindent 
\textbf{Resource theory of magic and stabilizer computation}---
The boundary between classically simulable and genuinely-advantageous quantum states was stated by the Gottesman--Knill theorem, which established that circuits restricted to Clifford gates and Pauli measurements can be classically simulated efficiently \cite{Gottesman1998, Aaronson2004}. 
The resource that lies across this boundary---\emph{magic}, or nonstabilizerness—measures the deviation of a state from the convex stabilizer~polytope. 
Bravyi and Kitaev showed that supplementing Clifford circuits with non-stabilizer ancillary states enables universal quantum computation, and that mixed states can be purified via magic state distillation under certain conditions~\cite{Bravyi2005}. 
This later motivated the formal resource theory of stabilizer computation, which introduced the concept of \emph{mana}---the total variation of the discrete Wigner representation---as a magic monotone and derived bounds on distillation efficiency~\cite{Veitch2014}. 
These bounds quantify the non-Clifford overhead in fault-tolerant architectures~\cite{Howard2017}.

\noindent 
\textbf{Quasidistribution techniques and error mitigation}---
There is a structural connection between magic and quasidistribution negativity: for odd-dimensional systems the mana is exactly the sum negativity of the discrete Wigner function, and~Delfosse \textit{et al.} extended this link to rebits~(quantum systems with states of real density-matrix representations in the computational basis) by proving a discrete Hudson theorem and establishing Wigner negativity as a computational resource for two-level systems~\cite{Delfosse2015}. 
More broadly, Raussendorf \textit{et al.} demonstrated that Wigner negativity and contextuality jointly serve as co-resources for qubit quantum computation~\cite{Raussendorf2017}. 
In the near-term setting, circuit-sampling scenarios framed in terms of quasidistributions directly demonstrate additional sampling cost due to the presence of~negativities. 
Reference~\cite{Temme2017} showed that the sampling overhead of probabilistic error cancellation---which inverts a noise channel by resampling circuits according to a quasidistribution decomposition---scales exponentially in the total variation of that distribution. 
That is, the weights corresponding to quasidistribution negativity enter as overhead contributions in error-mitigation protocols. 
On the other hand, noise-aware optimization of the decomposition basis may reduce this overhead~\cite{Piveteau2022}.

\noindent 
\textbf{Contextuality underlying magic and fault tolerance}---Howard \textit{et al.} proved that for odd-prime-dimensional quantum systems, contextuality is necessary \hj{\sout{and sufficient}} for universal quantum computation via magic state distillation~\cite{Howard2014}, and that all states outside the classically-simulable set~(possessing nonnegative quasidistributions) exhibit state-dependent~contextuality. 
This result was extended by Khu \textit{et al.}, who showed that contextuality is an intrinsic feature of error-correcting codes and fault-tolerant protocols~\cite{Khu2026}.

\noindent 
\textbf{Kirkwood--Dirac quasidistribution and quantum computational advantage}---Besides the popular methods involving the Wigner function, the generally complex Kirkwood--Dirac (KD) quasidistribution,
\begin{equation}
	\mathrm{KD}(\lambda,\mu) = \langle\lambda|\rho|\mu\rangle\langle\mu|\lambda\rangle
\end{equation}
for two arbitrary pure states $\ket{\mu}\bra{\mu}$ and $\ket{\lambda}\bra{\lambda}$ has emerged as a framework for diagnosing the quantumness of states across computation, metrology, and thermodynamics~\cite{Arvidsson2024,Lostaglio2023}. 
KD nonpositivity (the presence of negative real parts or nonzero imaginary parts) has been shown to be a necessary resource for quantum advantage in metrology, where anomalous weak values enabled by KD-negative states allow Fisher-information enhancements beyond classical limits~\cite{LupuGladstein2022}. 
Crucially, Thio \textit{et al.} demonstrated that KD nonpositivity is a resource monotone for quantum computation itself: in a rebit model of computation, all algorithms for which the KD distribution remains a proper positive probability measure throughout are efficiently classically simulable, establishing KD~nonpositivity as a necessary resource for computational quantum advantage~\cite{Thio2025}. 
Practical quantum circuits for measuring KD distributions were also proposed~\cite{Wagner2024, Wang2024}. 

\subsection{Quantum metrology}
One of the promising applications of quantum states is to achieve higher sensing precision beyond the classically allowed limit. 
In particular, parameter estimation enhanced by quantum-mechanical principles has opened the field of quantum metrology \cite{Giovannetti2006}, where quantum states of light have played a crucial role, as they are native to optical interferometer settings \cite{Barbieri2022}. 
One of the pioneering ideas was proposed by Carlton Caves \cite{Caves1981}, who proposed using the squeezed state of light in the interferometer to detect gravitational waves. 
This idea has been further developed into a practical approach to enhancing the sensitivity of the Laser Interferometer Gravitational-Wave Observatory (LIGO) \cite{Aasi2013} by leveraging quantum~states.

The classical limit of optical quantum metrology, the so-called standard quantum limit (SQL), is typically defined as the precision bound when using coherent states, i.e., classical states, as the probe state. 
This directly implies that non-classical states of light are a necessary resource for achieving quantum advantage in metrology.  
Not only squeezed states \cite{Matthews2016} but also various types of quantum light states, such as Fock states~\cite{Wolf2019, Deng2024, Rahman2025}, cat states~\cite{Munro2002, Joo2011,Joo2012, Tatsuta2019, Pan2023, Stammer2024, Zheng2025, Tatsuta2026}, NOON states \cite{Huver2008, Dowling2008, Lee2009, Llobet2020} and to a reasonable extent any Bosonic \cite{Oszmaniec2016} and non-Gaussian states \cite{Grochowski2025}, have been used as promising probe states for enhanced parameter estimation of phase \cite{Sanders1995} and displacement, reaching the Heisenberg limit~\cite{Ou1997}, the ultimate scaling limit allowed by quantum mechanics \cite{Giovannetti2004}.  
One may further look at the limitations of quantum metrology with quantum light in the presence of noise in \cite{Gorecki2022}. 
More detailed perspectives on quantumness related to quantum metrology can be found in \cite{Review_Tan2019, Berchera2019, Lawrie2019, Fadel2025}.

\section{Some open issues}
\label{sec:openissues}

\noindent 
\textbf{Negativity vs singularity and extremal states}---Equation~\eqref{eq:sqth_class_cond} shows that a squeezed thermal state turns quantum once $\nu/\mu> \bar{n}/(\bar{n}+1)$.
In that same regime, however, \eqref{eq:cf_sqth} yields no closed form for the $P$-function, because the Fourier transform diverges. 
The $P$-function is therefore expected to be highly singular.

Yet it is readily shown that the onset of quantumness signalled by an uncertainty below the classical floor---$\Delta x<1/2$ or $\Delta p<1/2$---already implies $P(\alpha)<0$ for squeezed thermal states. 
Whether the Sudarshan--Glauber $P$-function of such a state is \emph{negative} or \emph{singular} is thus left unclear.

This further opens up the question of which one is extremal \cite{Goldberg2020} between the negative-$P$ and singular-$P$ functions, at a given mean average energy. 
While \cite{Tan2020} sheds some light on the relation between the negativity and the singularity of the $P$-function, it is yet to be settled.

\noindent 
\textbf{Gaussian vs non-Gaussian $P$-function}---
Positive distribution functions can be simulated efficiently on classical hardware \cite{Mari2012, Veitch2013, Pashayan2015}, which makes negativity of a non-Gaussian Wigner function a clear marker of quantum advantage. 
Wigner negativity also certifies genuine non-Gaussianity, in the sense of lying outside the convex hull of Gaussian states.

The transformation \eqref{eq:p2w} relating the $P$- and Wigner functions preserves both Gaussianity and positivity, so every positive $P$-function yields a positive Wigner function. 
But \eqref{eq:p2w} is also a Gaussian smoothing, and smoothing can wash a negative $P$-function into a positive Wigner function. 
Attending to Wigner negativity alone therefore risks overlooking an entire class of genuinely non-Gaussian quantum states---states that may well matter for quantum computation. 
This raises, in turn, the question of genuine Gaussianity versus non-Gaussianity, and with it that of positive versus non-positive $P$-functions.

\noindent 
\textbf{The burden of optimization}---
Despite being mathematically rich and providing better understanding of foundational and pedagogical perspective of optical quantumness such as quantum to classical transition \cite{Lee2011, Sabapathy2016} or metrological advantages \cite{Hyukjoon2019, Ge2020}, most of the measures either require convex optimization \cite{Hillery1987a, Marian2002, Park2021, Bose2019} or are non-convex in nature \cite{Fu2022} or lack practical feasibility to a great extent \cite{Tan2017, Tan2020} as well as fail to address quantumness in general \cite{Bose2017}. 
This, in turn, necessitates an alternative description of the same that can be evaluated without optimization and is experimentally viable.

\noindent 
\textbf{Does quantumness imply contextuality?}---Einstein, Podolsky and Rosen argued that a quantum system carries definite values independently of what one chooses to measure \cite{EPR1935}. 
Later results overturned that presumption, in single-party as well as multiparty settings \cite{CochenSpecker1967}. 
This departure from the \emph{classical presumption of pre-existing values} is known as contextuality \cite{Spekkens2005}; it acts as a quantum resource \cite{Ferrie2011, Jeongwoo2024} and stands in exact correspondence with \emph{negativity of the Wigner function} \cite{Spekkens2008}. 
Wigner negativity is a quantum feature with no classical counterpart, yet optical quantumness is characterized more broadly and more deeply by the \emph{negativity or singularity} of the $P$-function. 
How contextuality relates to non-regularity of the $P$-function remains obscure, largely because of the mathematical intractability of the $P$-function itself.

\section{Concluding remarks}
\label{sec:conclusion}

In this review, we have presented a unified account of quantumness, emphasizing its mathematical characterization, physical interpretation, and operational relevance.  
Central to this framework is the Glauber–Sudarshan $P$-representation, which provides a necessary and sufficient criterion for distinguishing classical from quantum states of light.  Classical states admit a positive, well-defined probability distribution over coherent states, whereas quantumness arises from the breakdown of this description through negativity or singular behavior of the $P$-function.  This formulation establishes a direct connection between classical statistical optics and genuinely quantum features.

The difficulty is that the $P$-function is barely accessible experimentally: its singular structure for quantum states rules out direct reconstruction in most situations.  This has motivated the development of alternative approaches to quantify and detect quantumness.  These include Hilbert-space-based methods—such as distance measures, entropic quantifiers, and superposition-based criteria—as well as phase-space techniques based on quasidistributions and characteristic functions. Gaussian smoothing of the $P$-function, filtering at the level of characteristic functions, and the Wigner and Husimi distributions have proved especially practical in the laboratory. Despite their diversity, these approaches are closely interconnected and can often be interpreted within a unified framework based on phase-space transformations or equivalent quantum channels.

We have also examined how physical operations create and control quantumness.  Photon addition, quadrature squeezing, and Kerr nonlinear interactions all render a classical input quantum, whereas photon subtraction leaves classicality intact unless further resources are brought to bear. 
This asymmetry is intrinsic to quantum state engineering, and it offers concrete guidance for preparing quantum states in the laboratory.  
The way quantumness generation depends on parameters such as the squeezing strength illustrates, in addition, how continuously the classical and quantum regimes pass into one another.

From an operational standpoint, quantumness manifests through experimentally accessible signatures, including sub-Poissonian photon statistics, antibunching, quadrature squeezing, and higher-order correlations.  
Each detects quantumness without recourse to full state tomography, a decisive practical advantage.  Each also reflects the underlying structure of phase-space distributions and correlation functions, and so affords a complementary view of the state.

More recent works have established quantumness as a resource in a broader operational sense.  It is both necessary and sufficient for generating entanglement through linear optical transformations, which ties single-mode quantumness to multipartite quantum correlations. Similarly, in quantum metrology, quantum states enable enhanced parameter estimation beyond classical limits, as captured by the quantum Fisher information.  These connections highlight the central role of quantumness in quantum information processing and precision measurement.

Despite significant progress, several open challenges remain. 
A major issue is the absence of a universally applicable and experimentally efficient measure of quantumness capable of treating Gaussian and non-Gaussian states on equal footing.   Existing quantifiers—such as quantum depth, {quasidistribution} negativity, and robustness—often face limitations in generality, computational tractability, or operational interpretation.  Developing a consistent and scalable framework for quantifying quantumness remains an important direction for future research.

Extending these ideas to multimode and high-dimensional systems poses further difficulties, since mode correlations and coherence enter as additional structure. 
Quantumness in dynamical settings---open systems and noisy channels---is likewise unresolved, above all where quantumness and decoherence interact.

Looking ahead, advances in experimental techniques, including high-efficiency photodetection, homodyne tomography, and engineered nonlinear interactions, are expected to improve the generation and characterization of quantum states. At the same time, emerging approaches based on machine learning may offer efficient tools for quantumness detection and state reconstruction. From a theoretical perspective, further development of resource-theoretic frameworks incorporating operational constraints will be essential for connecting abstract measures with practical applications.

Optical quantumness is, in the end, both a fundamental feature of quantum theory and a working resource for quantum technologies. Understanding its structure, learning to quantify it, and putting it to use will remain central tasks for the field.

\section*{ACKNOWLEDGMENTS}
Over the years, we have benefited from questions, suggestions, criticism, and advice from many colleagues. Particular gratitude for help in various ways goes to Andrea Aiello, Gunnar Björk, Bob Boyd, Joe Eberly, Robert Fickler, Elisabeth Giacobino, Aaron Goldberg, Ebrahim Karimi, Antonio Z. Khoury, and Andrei Klimov. S.B. expresses his gratitude to Biswajit Chakraborty, RKMVERI, for several illuminating discussions.

This work was supported by the National Research Foundation of Korea (NRF) grant funded by the Korea government (MSIT) (No. RS-2024-00413957, No. RS-2024-00438415, No. RS-2023-NR076733, No. RS-2025-25464492, and No. RS-2026-25590066), the Institute of Information \& Communications Technology Planning \& Evaluation (IITP) grant funded by the Korea government (MSIT) (IITP-2025-RS-2020-II201606, IITP-2025-RS-2024-00437191, and RS-2025-02219034), and the Institute of Applied Physics at Seoul National University. H.K. is supported by the KIAS Individual Grant No. CG085302 at Korea Institute for Advanced Study. 
Y.S.T. acknowledges support from Global - Learning \& Academic research institution for Master’s·PhD students, and Postdocs(G-LAMP) Program of the National Research Foundation of Korea (NRF) grant funded by the Ministry of Education (RS-2026-25609336), and the faculty research fund of Sejong University in 2026.
The work of L.L.S.S.has been supported by the Agencia Estatal de Investigaci\'on (PID2025-172975NB-I00).


%
\end{document}